\documentclass[aps,pra,twocolumn,showpacs,superscriptaddress,groupedaddress]{revtex4}
\usepackage{bbm,amsmath,amssymb,amsbsy,graphicx,subfigure}

\usepackage{hyperref}
\usepackage[latin1]{inputenc}
\usepackage{txfonts}

\makeatletter
\DeclareRobustCommand{\element}[1]{\@element#1\@nil}
\def\@element#1#2\@nil{%
  #1%
  \if\relax#2\relax\else\MakeLowercase{#2}\fi}
\makeatother

\def\spc#1{\mathscr{#1}}

\def\comment#1{}

\def\grp#1{{\mathbf #1}}

\def\Reals{\ensuremath{\mathbb R}}
\def\Tr{\ensuremath{\mathrm{Tr}}}
\def\d{\ensuremath{\mathrm{d}}}
\def\spc#1{\mathcal{#1}}
\def\<{\ensuremath{\langle}}
\def\>{\ensuremath{\rangle}}

\newtheorem{Theo}{Theorem}

\def\Proof{\medskip\par\noindent{\bf Proof. }}
\def\qed{$\,\blacksquare$\par}

\newcommand{\barr}[1]{\overline{#1}}
\newcommand{\fc}{\barr{{\cal F}_c}}

\newcommand{\ket}[1]{|#1\rangle}
\newcommand{\bra}[1]{\langle#1|}

\begin{document}
\title{Quantum benchmarks for pure single-mode Gaussian states}

\date{January 7, 2014}

\author{Giulio Chiribella}
\email{gchiribella@mail.tsinghua.edu.cn}
\affiliation{Center for Quantum Information, Institute for Interdisciplinary
Information Sciences, Tsinghua University, Beijing, 100084, China}

 \author{Gerardo Adesso}
\email{gerardo.adesso@nottingham.ac.uk}
\affiliation{School of Mathematical Sciences, The University of Nottingham,
University Park, Nottingham NG7 2RD, United Kingdom}


\begin{abstract}
Teleportation and storage of continuous variable states of light and atoms are essential building blocks for the realization of large scale quantum networks. Rigorous validation of these implementations require identifying, and surpassing, benchmarks set by the most effective strategies attainable without the use of quantum resources. Such benchmarks have been established for special families of input states, like coherent states and particular subclasses of squeezed states. Here we solve the longstanding problem of defining quantum benchmarks for general pure Gaussian single-mode states with arbitrary phase, displacement, and squeezing, randomly sampled according to a realistic prior distribution.  As a special case, we show that the fidelity benchmark for teleporting squeezed states with totally random phase and squeezing degree is $1/2$, equal to the corresponding one for coherent states. We discuss the use of entangled resources to beat the benchmarks in  experiments.
\end{abstract} \maketitle


Quantum teleportation \cite{telep,vaidman,brakim} is the emblem of long-distance quantum communication \cite{repeaters} and provides a powerful primitive for quantum computing \cite{gotchu}. Similarly, quantum state storage \cite{storage} is a central ingredient for  quantum networks \cite{derid}. In the past two decades, the experimental progress in teleporting and storing quantum states realized on different physical systems has been impressive \cite{boschi,telezei,furuscience,teleatom1,teleatom2,naturusawa,telesqz1,telesqz2,memorysqz1,memorysqz2,memolaurat,remotememory,memsaab,memory10,memorypolzik,telepolzik,fernnp,fernnp13,furuscience12,zeil144}. Particularly groundbreaking are the demonstrations involving continuous variable (CV) systems \cite{brareview,book}, where states having an infinite-dimensional support, such as coherent and squeezed states, have been unconditionally teleported and stored between light modes and  atomic ensembles in virtually all possible combinations \cite{furuscience,memorypolzik,telepolzik,fernnp,fernnp13,cubic,hamrmp}. These experiments might be reckoned as  stepping stones for the {\it quantum internet} \cite{qinternet}.

 Ideally, teleportation and storage aim at the realization of a perfect identity channel between an unknown input state $\ket{\psi}_{\rm in}$,
 issued to the sender Alice, and the output state received by Bob. In principle, this is possible if Alice and Bob share a maximally entangled state, supplemented by classical communication \cite{telep,vaidman,brakim}. In practice, limitations on the available entanglement and technical imperfections lead to an output state $\rho_{\rm out}$ which is not, in general, a perfect replica of the input. It is then customary to quantify the success of the protocol in terms of the input-output fidelity \cite{uhlmann,jozsa} ${\cal F} = {}_{\rm in}\!\bra{\psi} \rho_{\rm out} \ket{\psi}_{\rm in}$, averaged over an ensemble $\Lambda=\{\ket{\psi}_{\rm in}, p_\psi\}$ of  possible input states, sampled according to a prior distribution known to Alice and Bob.
To assess whether the execution of transmission protocols takes advantage of genuine quantum resources, it is mandatory to establish benchmarks for the average fidelity \cite{brajmo}. A benchmark is given in terms of a threshold $\fc$, corresponding to the maximum average fidelity that can be reached without sharing any entanglement. Indeed, in a classical procedure Alice might just attempt to estimate $\ket{\psi}_{\rm in}$ through an appropriate measurement, and communicate the outcome to Bob, who could then prepare an output state based on such an outcome: this defines a ``measure-and-prepare'' strategy. For a given ensemble $\Lambda$, the \emph{classical fidelity threshold} (CFT)   $\fc$ amounts then to the highest average fidelity achievable by means of measure-and-prepare strategies.
If an actual implementation attains an average fidelity $\barr{{\cal F}_q}$ higher than $\fc$, then it is certified that no classical procedure could have reproduced the same results, and the quantumness of the implemented protocol is therefore validated. This is, in a sense  \cite{popqub,gispla,horobell,giulionew,bancalnew}, similar to observing a violation of Bell inequalities to testify the nonlocality of correlations in a quantum state \cite{bell,bellrev}.

In recent years, an intense activity has been devoted to devising appropriate benchmarks for teleportation and storage of relevant sets of input states \cite{popqub,benchqub,benchd,brajmo,hammerer,noi,owari,mariona,peterb,bancalnew}. In particular, if the ensemble $\Lambda $ contains arbitrary pure states of a $d$-dimensional system drawn according to a uniform distribution, then $\fc =2/(d+1)$ \cite{benchd}. In the limit of a CV system, $d \rightarrow \infty$, the CFT goes to zero, as it becomes impossible for Alice to guess a particular input state with a single measurement. However, for a quantum implementation  it is  meaningless  to assume that the laboratory source can produce arbitrary input states from an infinite-dimensional Hilbert space with nearly uniform probability distribution. To benchmark CV implementations one thus needs to restrict to  ensembles of input states that can be realistically prepared and are distributed according to  probability distributions with finite width.

In the  majority of CV protocols \cite{brareview}, {\it Gaussian states} have been employed as the preferred information carriers  \cite{pirandolareview}.
Gaussian states enjoy a privileged role as, on one hand,  their mathematical description only requires a finite number of variables (first and second moments of the canonical mode operators) \cite{ourreview}, and on the other, they represent the set of states which can be reliably engineered and manipulated in a multitude of laboratory setups \cite{book}. High-fidelity teleportation and storage architectures involving Gaussian states \cite{vaidman,brakim,furuscience,memorypolzik,telepolzik,cubic} can be scaled up to realize networks \cite{network,telepoppy,naturusawa} and hybrid teamworks \cite{teamwork}, and cascaded to build nonlinear gates for universal quantum computation \cite{cubic,pirandolareview}. The problem of benchmarking the  transmission of Gaussian states is thus of pressing relevance for quantum technology.

This problem has so far only witnessed partial solutions. Here and in the following, we shall focus on pure single-mode Gaussian states. Any such state  can be written as (we drop the subscript ``in'') \cite{ourreview,pirandolareview}
\begin{equation}\label{gauss}
\ket{\psi_{\alpha,s,\theta}} = \hat{D}(\alpha) \hat{S}(\xi) \ket{0}\,,
\end{equation}
where $\hat{D}(\alpha)=\exp(\alpha \hat{a}^\dagger-\alpha^{\ast} \hat{a})$ is the displacement operator, $\hat{S}(\xi) = \exp\big[\frac12(\xi{\hat{a}^{\dagger}}{}^2-\xi^{\ast} \hat{a}^2)\big]$ is the squeezing operator with $\xi = s e^{i \theta}$, $\hat{a}$ and $\hat{a}^\dagger$ are respectively the annihilation and creation operators obeying the relation $[\hat{a},\hat{a}^\dagger]=1$, and $\ket{k}$ denotes the $k^{\text{th}}$ Fock state, $\ket{0}$ being the vacuum. Pure single-mode Gaussian states are thus entirely specified by their displacement vector $\alpha \in \mathbb{C}$, their squeezing degree $s\in \mathbb{R}^+$, and their squeezing phase $\theta \in [0,2\pi]$.
A widely employed teleportation benchmark is available for the ensemble $\Lambda_C$ of input {\it coherent states} \cite{brajmo,furuscience,hammerer}, for which  $s,\theta=0$ and the displacement $\alpha$ is sampled according to a Gaussian distribution $p^C_\lambda(\alpha)=\frac{\lambda}{\pi}e^{-\lambda |\alpha|^2}$ of width $\lambda^{-1}$. In this case, the CFT reads  \cite{hammerer}
\begin{equation}\label{cbench}
{\fc}^C(\lambda) = \frac{1+\lambda}{2+\lambda}\,,
\end{equation}
converging to $\lim_{\lambda \rightarrow 0} {\fc}^C(\lambda)=\frac12$ in the limit of infinite width. More recently, benchmarks were obtained for particular subensembles of {\it squeezed states} \cite{noi,owari,mariona}, specifically either for known $s$ and totally unknown $\alpha,\theta$ \cite{owari}, or for totally unknown $s$ with $\alpha,\theta=0$ \cite{noi,noinote}.  However, up to date a fundamental question has remained unanswered in CV quantum communication: {\it What is the general benchmark for teleportation and storage of arbitrary pure single-mode Gaussian states?}

In this Letter we solve this longstanding open problem. We build on a recent method for the evaluation of quantum benchmarks proposed in Ref.~\cite{giulionew}, and develop group-theoretical techniques to calculate the CFT for the following two classes of input single-mode states: (a) the ensemble $\Lambda_S $, containing pure Gaussian squeezed states with no displacement ($\alpha=0$), totally random phase $\theta$, and unknown squeezing degree $s$ drawn according to a realistic distribution with width $\beta^{-1}$; (b) the ensemble $\Lambda_G$, containing arbitrary pure Gaussian states with totally random phase $\theta$ and $\alpha,s$ drawn according to a joint distribution with finite widths $\lambda^{-1}, \beta^{-1}$, respectively. By properly selecting the prior distributions, we obtain analytical results for the benchmarks, which eventually take the following simple and intuitive form:
\begin{subequations}\label{hobgoblin}
\begin{align}
{\fc}^S(\beta) &= \frac{1+\beta}{2+\beta}\,; \label{sbench}\\
{\fc}^G(\lambda,\beta) &= \frac{(1+\lambda)(1+\beta)}{(2+\lambda)(2+\beta)}\,.\label{gbench}
\end{align}
\end{subequations}
These benchmarks are \emph{probabilistic}  \cite{giulionew}: they give the maximum of the fidelity over arbitrary measure-and-prepare strategies,  even including probabilistic strategies based on post-selection of some measurement outcomes. By definition, probabilistic benchmarks are stronger than deterministic ones: beating a probabilistic benchmark means having an implementation whose performance cannot be achieved classically, even with a small probability of success.

Case (a) shows that for input squeezed states with totally unknown complex squeezing $\xi$, the benchmark reaches $\lim_{\beta \rightarrow 0}{\fc}^S(\beta)=\frac12$ just like the case of coherent states; we provide a nearly optimal measure-and-prepare deterministic strategy which saturates the benchmark of Eq.~(\ref{sbench}) for $\beta \gg 0$. On the other hand, the general result of case (b) encompasses the previous partial findings providing an elegant and useful prescription to validate experiments involving transmission of Gaussian states, with input distribution widths $\lambda^{-1},\beta^{-1}$ tunable  depending on the  capabilities of actual implementations.

{\it Mathematical formulation of quantum benchmarks.}---
Suppose that Alice and Bob want to teleport/store a state chosen at random from an ensemble $\{  |\varphi_x\>, p_x\}_{x\in\mathsf X}$ using a measure-and-prepare strategy, where Alice measures the input state with a positive operator-valued measure (POVM) $\{P_y\}_{y\in\mathsf Y}$ and, conditionally on outcome $y$, Bob prepares an output state $\rho_y'$.   In a probabilistic strategy, Alice and Bob have the extra freedom to discard some of the measurement outcomes and to produce an output state only when the outcome $y$ belongs to a set of favourable outcomes $\mathsf Y_{\text{yes}}$.   The fidelity of their  strategy is
\begin{align}\label{fid}
\overline{\mathcal F}  =   \sum_{x\in\mathsf X} \sum_{y\in\mathsf Y_{\text{yes}}}  p (x|\text{yes}) ~  p(y|x,\text{yes})  ~  \<  \varphi_x|  \rho_y  | \varphi_x\> ,
\end{align}
where $p(x|\text{yes})$ is the conditional probability of having the state $|\varphi_x\>$ given that a favourable outcome  was observed and $p(y|x,\text{yes})  :=  \<  \varphi_x|   P_y  |\varphi_x\>  /  \sum_{y'\in \mathsf Y_{\text{yes}}}      \<  \varphi_x|   P_{y'}  |\varphi_x\>$.  Then the CFT is the supremum of Eq.~(\ref{fid}) over all possible measure-and-prepare strategies. Using a result of \cite{giulionew}, we have 
\begin{align}\label{general}
\fc  =   \left\|   \left(I\otimes \tau^{-1/2}\right) \rho    \left(I\otimes \tau^{-1/2}\right) \right\|_\times,
\end{align}
where $\tau =  \sum_x  p_x  |\varphi_x\>\<\varphi_x|$ is the average state of the ensemble, $\rho =  \sum_x p_x  |\varphi_x\>\<\varphi_x|  \otimes |\varphi_x\>\<\varphi_x| $, and, for a positive operator $A$,  $\| A  \|_\times  =   \sup_{  \|  \varphi   \|  =  \|  \psi \| = 1}  \< \varphi|  \<\psi|  A  | \varphi\>|\psi\>$.

\begin{figure}[t]
\subfigure[]{
\includegraphics[height=2.4cm]{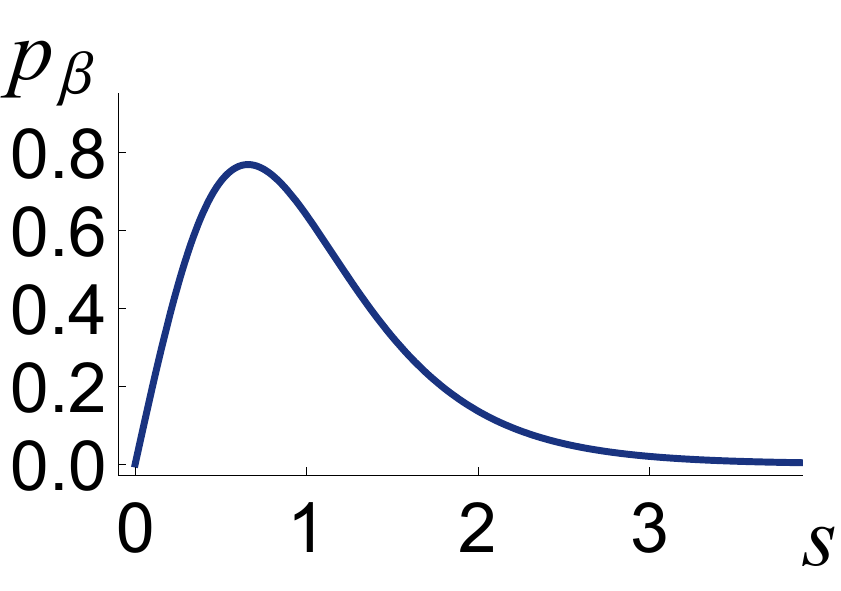}\label{pba}}\hspace*{.3cm}
\subfigure[]{
\includegraphics[height=2.4cm]{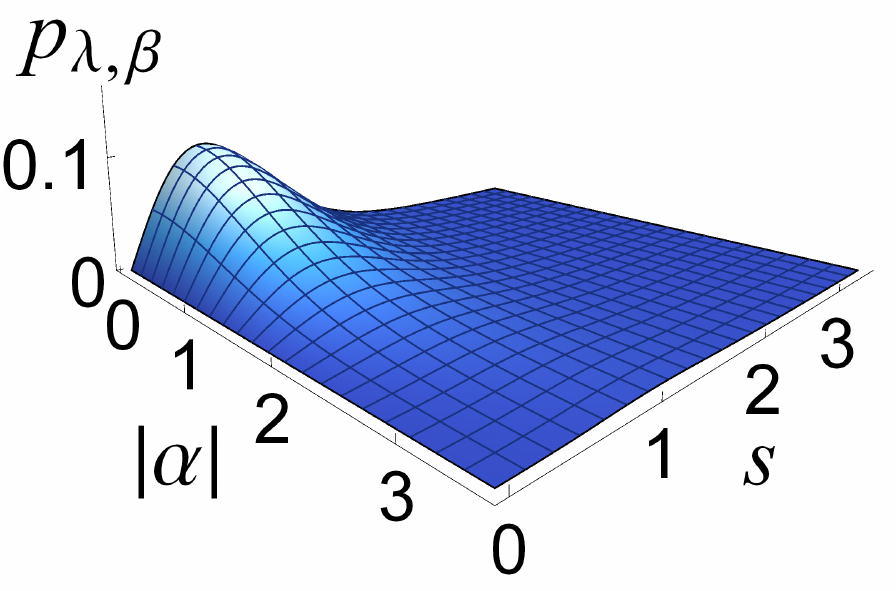}\label{pbb}}
\caption{  (Color online) Marginal probability distributions for (a) the subset of input squeezed states with arbitrary squeezing degree $s$ and arbitrary phase $\theta$, and (b) the complete set of  input Gaussian states with arbitrary displacement $\alpha$, arbitrary squeezing degree $s$, and arbitrary random phase $\theta$. The plots show the marginal distributions after integrating over $\theta$, having  set $\beta=2$, $\lambda=\frac12$.}
\label{figpigs}
\end{figure}

{\it Case (a): Benchmark for arbitrary squeezed states.}---
We consider the ensemble $\Lambda_S$ of squeezed vacuum states $|\xi \> \equiv \ket{\psi_{0,s,\theta}} $ with arbitrary complex squeezing [see Eq.~(\ref{gauss})] distributed
according to the prior
\begin{equation}\label{pbeta}
p^S_\beta(\xi)  = \frac{p_\beta(s)  }{2\pi}\,,\quad  \mbox{with} \ \ p_\beta(s) =  \frac{  \beta     \sinh s }{ (\cosh s)^{\beta + 1}} \,,
\end{equation}
where $\beta^{-1} >0$ regulates the width of the squeezing distribution,
while the phase $\theta$ is  uniformly distributed, which is natural for CV experiments \cite{mariona}. The marginal prior $p_\beta(s)$ is plotted in Fig.~\ref{figpigs}(a).  For squeezed states, the prior  $p_\beta^S(\xi)$ is the analogue of the Gaussian $p_\lambda^C (\alpha)$ for coherent states: indeed, the Gaussian can be expressed as $p_\lambda^C (\alpha) \d^2 \alpha= |\<  0 |  \alpha\>|^{2\lambda}  \d^2  \alpha/\pi$, where the measure $\d^2 \alpha$ is invariant under the action of displacements, while $p_\beta^S(\xi)$ can be expressed as $p_\beta^S(\xi)~\d s \d \theta  = | \<  0  |  \xi\>|^{2(2+\beta)}   \mu(\d^2 \xi)$, where the measure $\mu(\d^2\xi)  =  \sinh s  \cosh s ~\d s \d \theta/(2\pi)$ is  invariant  under the action of the squeezing transformations.  For integer $\beta$, the prior $p_{\beta}^S(\xi)$ can be generated by preparing $2+\beta$ modes in the vacuum and  performing the optimal measurement for the estimation of squeezing 
\cite{hayashi,epaps}.

Using Eq.~(\ref{general}), the CFT  can be written as
$\fc^S(\beta) =  \| A_\beta \|_\times, \,  A_\beta  =  (  I  \otimes \tau_\beta^{-1/2})  \rho_\beta  ( I\otimes \tau_\beta^{-1/2} )$
where  $\tau_\beta =  \int  \d s \d \theta ~ p^S_\beta(\xi)    |\xi\>\< \xi  | $ and $\rho_\beta  =  \int \d s \d \theta ~ p^S_\beta(\xi) ~   |  \xi\>\< \xi|  \otimes |\xi\>\<  \xi|$.
To obtain the benchmark announced in Eq.~(\ref{sbench}), we compute explicitly the states $\tau_\beta$ and $\rho_\beta$ and show that the eigenvalues of  $A_\beta $ are all equal to $\frac{1+\beta}{2+\beta}$ \cite{epaps}. Observing that  $  \frac{1+\beta}{2+\beta}  = \< 0  |  \<0  | A_\beta  |0\>|0\> $, we then get
 $\|  A_\beta \|_\times    =\frac{1+\beta}{2+\beta}$,
thus concluding the proof of Eq.~(\ref{sbench}).
This benchmark allows one to certify the quantumness of experiments involving teleportation and storage of squeezed states with arbitrary amount of squeezing and arbitrary phase \cite{telesqz1,telesqz2,memorysqz1,memorysqz2,fernnp}, bypassing the limitations of \cite{owari,noi}.

\begin{figure}[t]
\includegraphics[width=7cm]{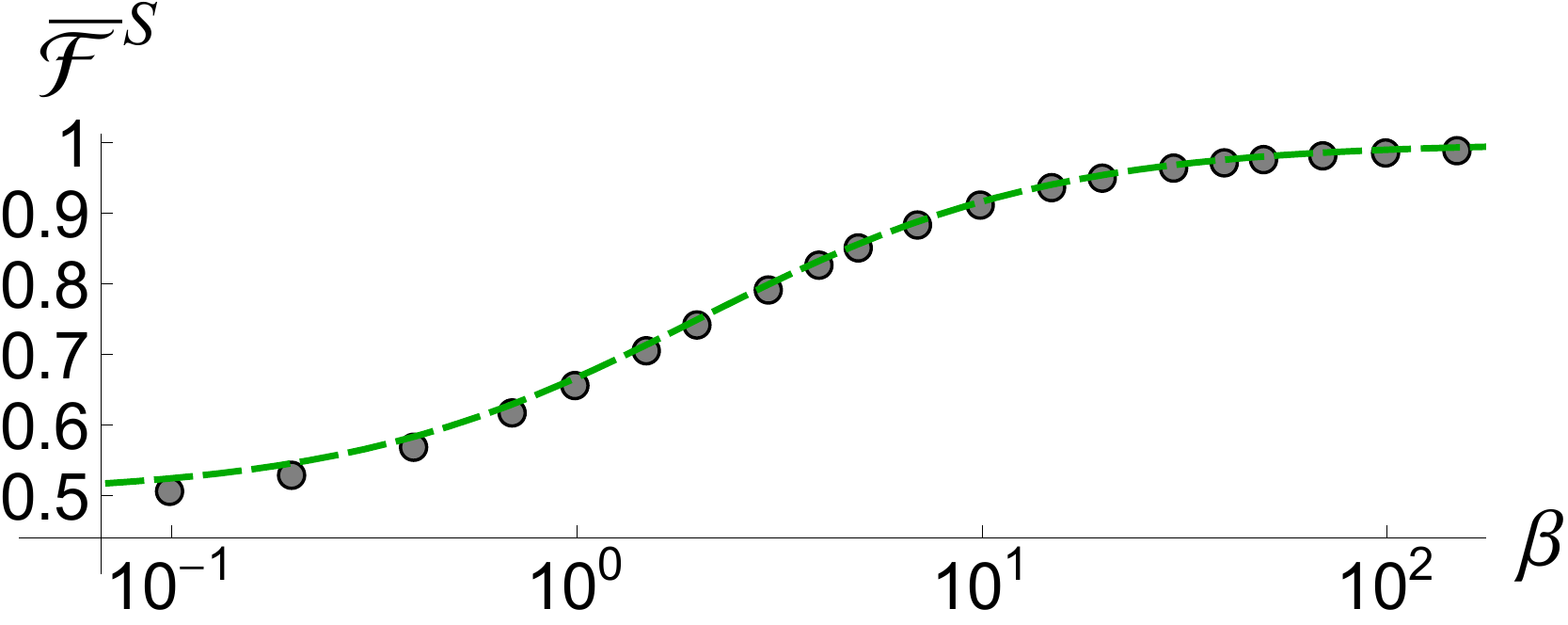}
\caption{  (Color online) Average classical fidelities for input squeezed states versus the distribution parameter $\beta$. The dots correspond to the fidelity $\barr{{\cal F}_{sr}}^S$ for the best square-root measurement, while the dashed line depicts the optimal probabilistic CFT $\fc^S$.\label{figsqrt}}
\end{figure}

We highlight the similarity of our result to the case of input coherent states \cite{hammerer}. In that case, the probabilistic benchmark of Eq.~(\ref{general}) coincides with the maximum over deterministic strategies, given by Eq.~(\ref{cbench}) \cite{giulionew}. Precisely, the CFT of Eq.~(\ref{cbench}) is achievable with  heterodyne detection and repreparation of coherent states \cite{brajmo,hammerer}.   Since the heterodyne detection can be interpreted as a square-root measurement \cite{square,srm} for a suitable Gaussian prior,  in the case of squeezed states it is natural to wonder  whether a deterministic square-root measurement strategy suffices to saturate the probabilistic CFT given by Eq.~(\ref{sbench}). For an ensemble of the form $\{|\xi\>,  p^S_\eta  (\xi)\} $ the square-root measurement has POVM elements $P_\eta(\xi)   =   p^S_\eta(\xi)  \tau_\eta^{-1/2}     |\xi\>\< \xi|            \tau_\eta^{-1/2}$ (here we allow $\eta$ to be different from $\beta$). Performing the square-root measurement and repreparing the state $|\xi\>$ conditional on outcome $\xi$ gives the average fidelity
$ \barr{{\cal F}_{sr}}^S (\beta,\eta)  =   
 \nonumber  \frac{\beta}{\beta +2}  \frac{ \eta+ 1}{\eta+2}
  \sum_{k=0}^\infty    {\left|  \sum_{n=0}^k  {k-n -\frac 12 \choose  k-n}    \sqrt{ {   n-\frac 12\choose n  } {  \frac {\eta + 1} 2  +n \choose  n  } } \right|^2}/\left[
 { {  \frac {  \beta + 2}2  +  k  \choose k}   {  \frac {  \eta + 2}2  +  k  \choose k}  }\right]$  \cite{epaps}, where we are using the notation $ {x  \choose k } =\frac{  x (x-1)  \dots (x-k+1)}{k!}$ for a general $x\in\Reals$. 
 In Fig.~\ref{figsqrt} we compare $\sup_\eta\barr{{\cal F}_{sr}}^S (\beta,\eta)$, maximized numerically over $\eta$, with the CFT $\fc^S$ of Eq.~(\ref{sbench}), for a range of values of $\beta$. We find that the square-root measurement is a nearly optimal classical strategy, which reaches the CFT asymptotically for large values of $\beta$, when the input squeezing distribution becomes more and more peaked.

{\it Case (b): Benchmark for general Gaussian  states.}---
Consider now the ensemble $\Lambda_G$ of  arbitrary pure Gaussian states $\ket{\alpha,\xi} \equiv\ket{\psi_{\alpha,s,\theta}}$ [Eq.~(\ref{gauss})] distributed according to the prior
\begin{eqnarray}
\hspace*{-1cm}p_{\lambda,\beta}^G(\alpha,s,\theta)&=&\frac{\lambda \beta}{2 \pi^2}        \frac{ e^{-  \lambda|\alpha|^2   + \lambda  {\rm Re} (  e^{-i\theta} \alpha^2)  \tanh s    }   \sinh s}{(\cosh s)^{\beta + 2}}  \,.\label{plambdabeta}
\end{eqnarray}
We note that in this case the prior can be written as $p^G_{\lambda,\beta}  (\xi)  \propto   |\<  0|  \lambda \alpha, \xi\>|^2   |\<  0  |  \xi\>|^{2(4+\beta)}  \nu(\d^2\alpha,\d^2 \xi) $ where $  \nu(\d^2\alpha,\d^2 \xi)  =  \d^2\alpha  \sinh s (\cosh s)^3 \d s \d \theta$ is the invariant measure under the joint action of displacement and squeezing. For integer $\beta$, the prior can be generated by performing an optimal measurement of squeezing and displacement on $5+\beta$ modes prepared in the vacuum \cite{epaps}. The marginals of this prior correctly reproduce the previous subcases, namely
the distribution of Eq.~(\ref{pbeta}) for the squeezing,   $\int \d^2\alpha\ p^G_{\lambda,\beta}(\alpha,s,\theta) = p^S_\beta(\xi)$, and the Gaussian distribution of \cite{hammerer} for the displacement, $\lim_{\beta \rightarrow \infty} \int \d^2\xi\ p^G_{\lambda,\beta}(\alpha,s,\theta) = p^C_\lambda(\alpha)$. The marginal probability distribution after integrating over the phase $\theta$, $p_{\lambda,\beta}(\alpha,s) = \int_0^{2\pi} \d \theta\ p^G_{\lambda,\beta}(\alpha,s,\theta) =
 \pi^{-1}  \lambda \beta  e^{-\lambda |\alpha|^2}  \sinh s (\cosh s)^{-\beta
   -2} I_0\big[\lambda|\alpha|^2   \tanh s\big]$, where $I_0$ is a modified Bessel function \cite{borntobe}, is plotted in Fig.~\ref{figpigs}(b).

\begin{figure}[tb]
\subfigure[]{\includegraphics[height=3.9cm]{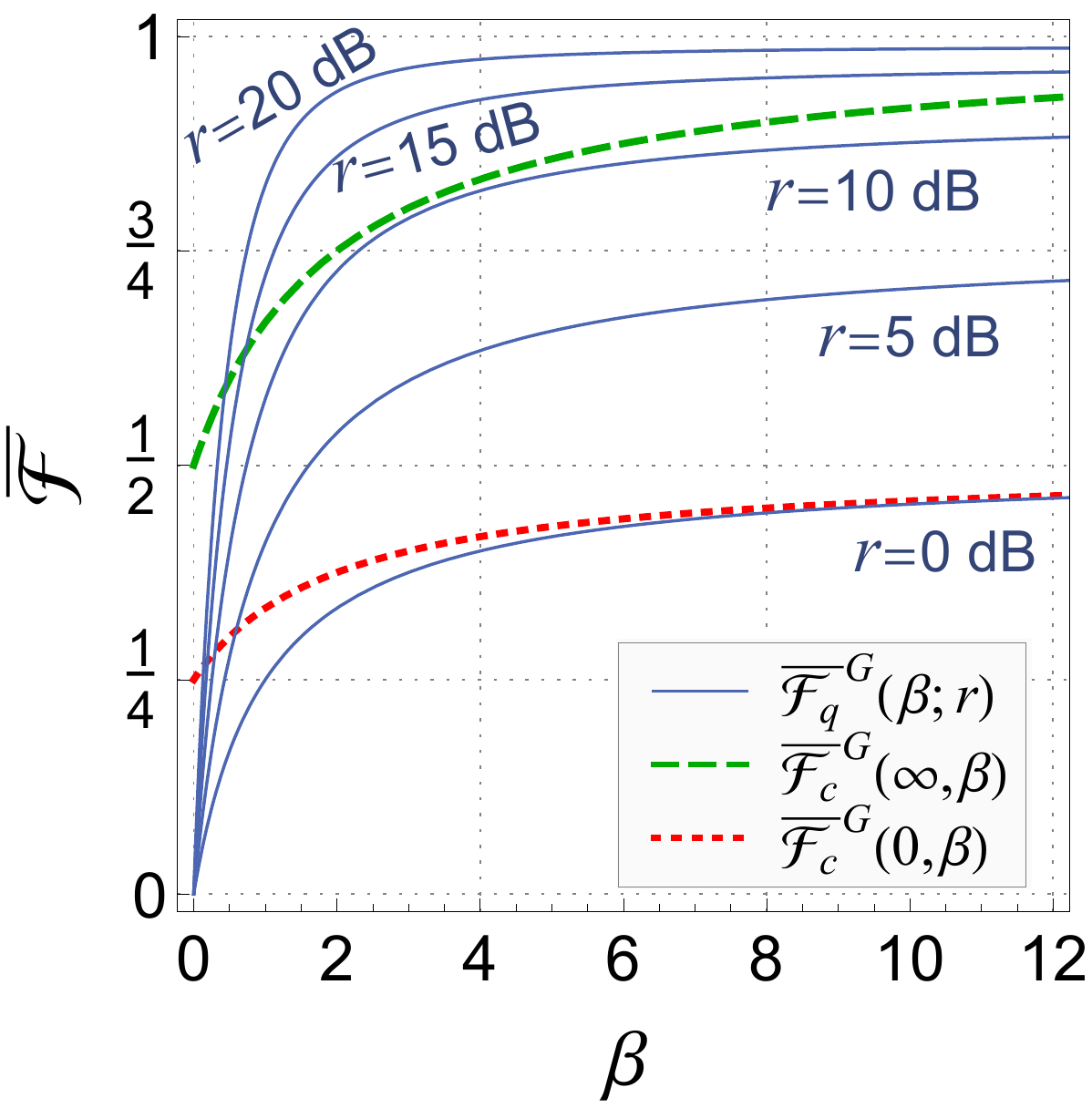}}\hspace*{.1cm}
\subfigure[]{\includegraphics[height=3.9cm]{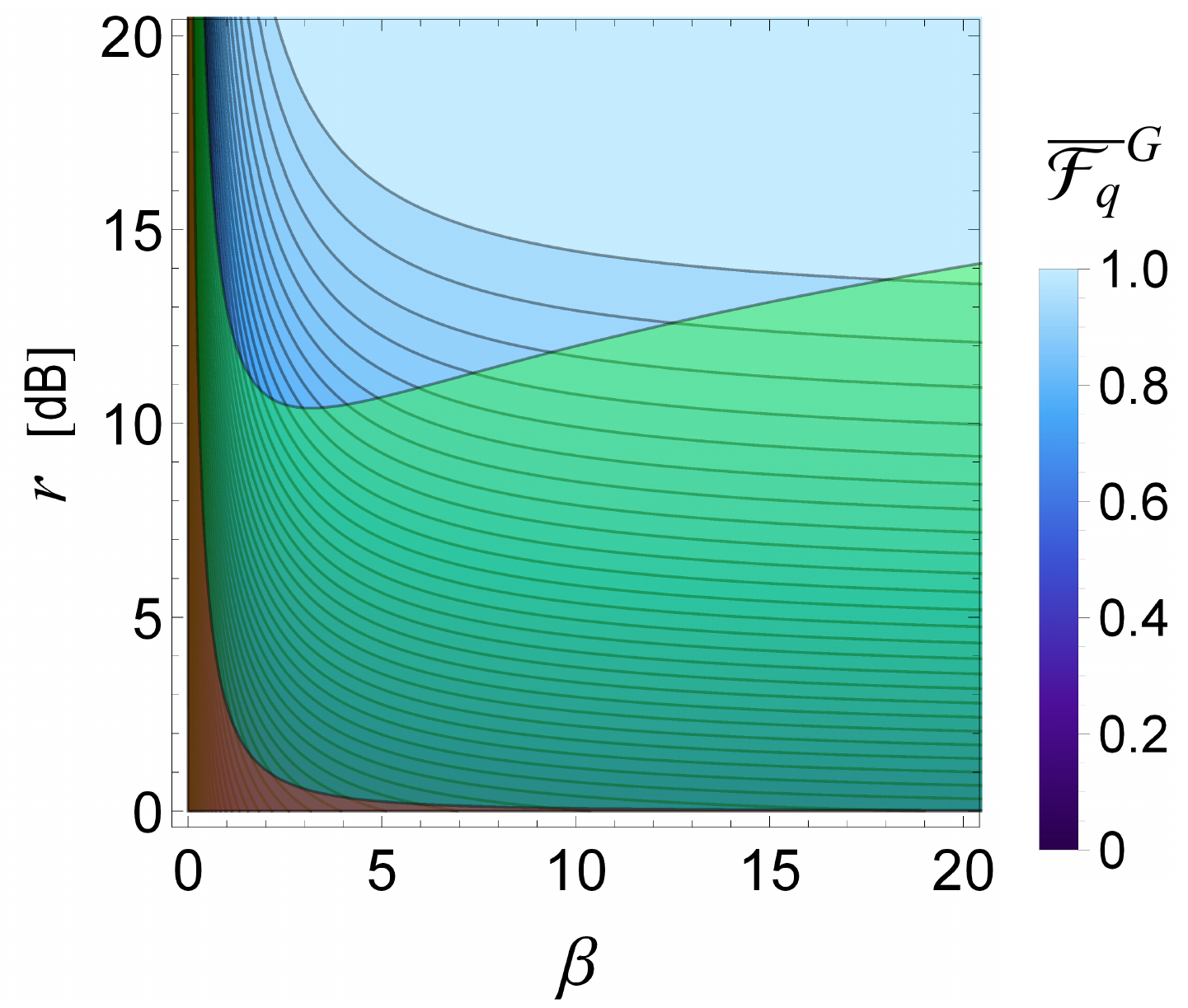}}  
    \caption{
    (Color online) Performance of the  CV quantum teleportation protocol  for general input Gaussian states $\ket{\psi_{\alpha,s,\theta}} $ using a two-mode squeezed entangled resource with squeezing $r$. (a) Plot of the quantum teleportation fidelity $\barr{{\cal F}_q}^G(\beta;r)$, averaged over the input set $\Lambda_G$ according to a prior distribution $p_{\lambda,\beta}$,  against the benchmark $\fc^G(\lambda,\beta)$ for $\lambda \rightarrow 0$ (dotted red) and $\lambda \rightarrow \infty$ (dashed green).  (b) Contour plot of $\barr{{\cal F}_q}^G(\beta;r)$ as a function of $\beta$ and $r$;  the lower (red) and upper (green) shadings correspond to parameter regions where the quantum fidelity does not beat the benchmark $\fc^G(0,\beta)$ and $\fc^G(\infty,\beta)$, respectively. } \label{fient}
\end{figure}

To compute the benchmark, we observe that the pure Gaussian states of Eq.~(\ref{gauss}) are instances of the generalized coherent states  introduced by Gilmore and Perelomov for arbitrary Lie groups \cite{gilmore,perelomov,perelomovbook,ali}.
Here we consider Gilmore-Perelomov coherent states of the form  $|\varphi_g\>=  \hat{U}_g  |\varphi\>$, where   $\hat{U}: g\mapsto \hat{U}_g$  is an irreducible representation of a Lie group $\mathsf G$ and $|\varphi\>$ is a lowest weight vector for the representation  $\hat{U}: g\mapsto \hat{U}_g$.
This general setting includes the cases of coherent and squeezed states, and the present case of pure Gaussian states, where the group is the Jacobi group, the group element $g$ is the pair $g =  (\alpha,\xi)$, $\hat{U}_g \equiv \hat{D}(\alpha) \hat{S} (\xi)$, and $\ket{\varphi} \equiv \ket{0}$  \cite{berceanu}. In the Supplemental Material \cite{epaps}, we solve the  benchmark problem for arbitrary sets of Gilmore-Perelomov coherent states,  randomly drawn with a prior probability of the form $p_\gamma(g) \d g   \propto \, |\<   \varphi_{\gamma}   |\varphi_{\gamma,g}\>   |^2\,  \d g$,
where $\d g$ is the invariant measure on the group and $ |\varphi_{\gamma,g}\>  =  \hat U^\gamma_g  |\varphi_\gamma\>$ is the Gilmore-Perelomov coherent state for a given irreducible representation $\hat{U}^\gamma  :  g \mapsto \hat{U}_g^\gamma$.
Our key result is  a powerful formula for   the probabilistic CFT for Gilmore-Perelomov coherent states,  given by  \cite{epaps}
\begin{equation}\label{hobPerel}
\fc(\gamma)  =
\frac  {  \int \d g~  p_\gamma  (g ) ~  | \<  \varphi|  \varphi_g\> |^4 }  {  \int \d g  ~ p_\gamma  (g ) ~   |\<  \varphi|  \varphi_g\> |^2  } .
\end{equation}
Using this general expression in the cases of coherent and squeezed states it is immediate to retrieve the benchmarks of Eqs.~(\ref{cbench}) and (\ref{sbench}).   We now  use this  result to find the benchmark for the transmission of arbitrary input Gaussian states with prior distribution given by Eq.~(\ref{plambdabeta}), which now reads
\begin{equation}\label{cftg}
\fc^G(\lambda,\beta) = \frac
{\int \d^2\alpha\,\d s \,\d\theta\ p^G_{\lambda,\beta}(\alpha,s,\theta)    | \<  0  |  \psi_{\alpha,s,\theta} \> |^4  }
{\int \d^2\alpha\,\d s \,\d\theta\ p^G_{\lambda,\beta}(\alpha,s,\theta)       | \<  0  |  \psi_{\alpha,s,\theta} \> |^2 }\,.
\end{equation}
The integrals can be evaluated analytically \cite{epaps}. The final result yields the general benchmark announced in
Eq.~(\ref{gbench}), which is the main contribution of this Letter.
Notice how  the previous partial findings are contained in this result. For coherent states, $\lim_{\beta \rightarrow \infty} \fc^G(\lambda,\beta) = \fc^C(\lambda)$; for squeezed states, $\lim_{\lambda \rightarrow \infty} \fc^G(\lambda,\beta) = \fc^S(\beta)$. The benchmark for teleporting Gaussian states in the limit of completely random $\alpha, s, \theta$ is finally established to be $\lim_{\lambda,\beta \rightarrow 0} \fc^G(\lambda,\beta) = \frac14$.

   {\it Discussion.}---
We now investigate how well an actual implementation of quantum teleportation can fare against the benchmarks derived above. We focus on the conventional Braunstein--Kimble CV quantum teleportation protocol \cite{brakim} using as a resource a Gaussian two-mode squeezed vacuum state with squeezing $r$, $\ket{\phi}_{AB} = (\cosh r)^{-1} \sum_{k=0}^{\infty} (\tanh r)^k \ket{k}_A \ket{k}_B$, also known as a twin-beam. We assume that the input is an arbitrary pure single-mode Gaussian state $\ket{\psi_{\alpha,s,\theta}}$, Eq.~(\ref{gauss}), drawn according to the probability distribution of Eq.~(\ref{plambdabeta}).  The output state received by Bob will be a Gaussian mixed state whose fidelity with the input can be written as \cite{noi}
${\cal F}_q(s;r) = \{2 e^{-2r} [\cosh (2r) + \cosh (2s)]\}^{-\frac12}$. Notice that  it depends neither on the phase $\theta$ nor the displacement $\alpha$ by construction of the CV protocol \cite{brakim} (for unit gain \cite{vloko,cubic}). Averaging this over the input set $\Lambda_G$ we get the average quantum teleportation fidelity
\begin{eqnarray}\label{qft}
\barr{{\cal F}_q}^G(\beta;r) &=&  \begin{array}{c}\int \d\theta\,\d^2\alpha\,\d s\ p^G_{\lambda,\beta}(\alpha,s,\theta) {\cal F}_q(s;r)\end{array}  \\
&=&\begin{array}{c}\frac{\beta}{2 \beta +2}  e^r \  _2F_1\left(\frac{1}{2},\frac{\beta +1}{2};\frac{\beta
   +3}{2};-\sinh ^2(r)\right)\end{array} \nonumber ,
\end{eqnarray}
where $_2F_1$ is a hypergeometric function \cite{borntobe}.
The average quantum fidelity is obviously independent of $\lambda$, i.e., in particular, it is the same for the ensemble of all Gaussian states $\Lambda_G$ and for the ensemble of squeezed states $\Lambda_S$.
In Fig.~\ref{fient}, we compare $\barr{{\cal F}_q}^G(\beta;r)$ with the CFT $\fc^G(\lambda,\beta)$, in particular with the case $\lambda\rightarrow 0$ (totally random displacement) and with the case $\lambda \rightarrow \infty$ (undisplaced squeezed states, whose CFT reduces to $\fc^S(\beta)$). In the latter case, we see that the shared entangled state needs to have a squeezing $r$ above $10$ dB, which is at the edge of current technology \cite{schnabel,schnabel13}, in order to beat the benchmark for the ensemble $\Lambda_S$. For general input Gaussian states in $\Lambda_G$ with random displacement, squeezing, and phase, less resources are instead needed to surpass the CFT of Eq.~(\ref{gbench}), especially if the input squeezing distribution is not too broad ($\beta \gg 0$), which is the realistic situation in experimental implementations (where e.g.~$s$ can fluctuate around a set value which depends on the specifics of the nonlinear crystal used for optical parametric amplification \cite{book,cubic}).   For the case of coherent input states with totally random displacement ($\lambda \rightarrow 0, \beta \rightarrow \infty$), the CFT converges to $\frac12$ and we recover the known result that any $r>0$ is enough to beat the corresponding benchmark  \cite{brakim,brajmo,furuscience,hammerer,telepoppy}.

Summarizing, we have derived exact analytical quantum benchmarks for teleportation and storage of arbitrary pure single-mode Gaussian states, which can be readily employed to validate current and future implementations. The mathematical techniques developed here to obtain the presented results are of immediate usefulness to analyze a much  larger class of problems, such as the determination of benchmarks for cloning, amplification \cite{giulionew} and other protocols involving multimode Gaussian states and other classes of Gilmore-Perelomov coherent states, including finite-dimensional states. We will explore  these topics in forthcoming publications.

{\it Acknowledgments.}---This work was supported by the Tsinghua--Nottingham  Teaching and Research Fund. GC is supported by the National Basic Research Program of China (973) 2011CBA00300 (2011CBA00301), by the National Natural Science Foundation of China (Grants 11350110207, 61033001,  61061130540), and by the 1000 Youth Fellowship Program of China. GC thanks S.~Berceanu for useful discussions on the Jacobi group. GA thanks N.~G.~de~Almeida for discussions and the Brazilian agency CAPES (Pesquisador Visitante Especial-Grant No. 108/2012) for financial support.


\begin{thebibliography}{68}
\expandafter\ifx\csname natexlab\endcsname\relax\def\natexlab#1{#1}\fi
\expandafter\ifx\csname bibnamefont\endcsname\relax
  \def\bibnamefont#1{#1}\fi
\expandafter\ifx\csname bibfnamefont\endcsname\relax
  \def\bibfnamefont#1{#1}\fi
\expandafter\ifx\csname citenamefont\endcsname\relax
  \def\citenamefont#1{#1}\fi
\expandafter\ifx\csname url\endcsname\relax
  \def\url#1{\texttt{#1}}\fi
\expandafter\ifx\csname urlprefix\endcsname\relax\def\urlprefix{URL }\fi
\providecommand{\bibinfo}[2]{#2}
\providecommand{\eprint}[2][]{\url{#2}}

\bibitem[{\citenamefont{Bennett et~al.}(1993)\citenamefont{Bennett, Brassard,
  Cr\'epeau, Jozsa, Peres, and Wootters}}]{telep}
\bibinfo{author}{\bibfnamefont{C.~H.} \bibnamefont{Bennett}},
  \bibinfo{author}{\bibfnamefont{G.}~\bibnamefont{Brassard}},
  \bibinfo{author}{\bibfnamefont{C.}~\bibnamefont{Cr\'epeau}},
  \bibinfo{author}{\bibfnamefont{R.}~\bibnamefont{Jozsa}},
  \bibinfo{author}{\bibfnamefont{A.}~\bibnamefont{Peres}}, \bibnamefont{and}
  \bibinfo{author}{\bibfnamefont{W.~K.} \bibnamefont{Wootters}},
  \bibinfo{journal}{Phys. Rev. Lett.} \textbf{\bibinfo{volume}{70}},
  \bibinfo{pages}{1895} (\bibinfo{year}{1993}).

\bibitem[{\citenamefont{Vaidman}(1994)}]{vaidman}
\bibinfo{author}{\bibfnamefont{L.}~\bibnamefont{Vaidman}},
  \bibinfo{journal}{Phys. Rev. A} \textbf{\bibinfo{volume}{49}},
  \bibinfo{pages}{1473} (\bibinfo{year}{1994}).

\bibitem[{\citenamefont{Braunstein and Kimble}(1998)}]{brakim}
\bibinfo{author}{\bibfnamefont{S.~L.} \bibnamefont{Braunstein}}
  \bibnamefont{and} \bibinfo{author}{\bibfnamefont{H.~J.}
  \bibnamefont{Kimble}}, \bibinfo{journal}{Phys. Rev. Lett.}
  \textbf{\bibinfo{volume}{80}}, \bibinfo{pages}{869} (\bibinfo{year}{1998}).

\bibitem[{\citenamefont{Briegel et~al.}(1998)\citenamefont{Briegel, D\"ur,
  Cirac, and Zoller}}]{repeaters}
\bibinfo{author}{\bibfnamefont{H.-J.} \bibnamefont{Briegel}},
  \bibinfo{author}{\bibfnamefont{W.}~\bibnamefont{D\"ur}},
  \bibinfo{author}{\bibfnamefont{J.~I.} \bibnamefont{Cirac}}, \bibnamefont{and}
  \bibinfo{author}{\bibfnamefont{P.}~\bibnamefont{Zoller}},
  \bibinfo{journal}{Phys. Rev. Lett.} \textbf{\bibinfo{volume}{81}},
  \bibinfo{pages}{5932} (\bibinfo{year}{1998}).

\bibitem[{\citenamefont{Gottesman and Chuang}(1999)}]{gotchu}
\bibinfo{author}{\bibfnamefont{D.}~\bibnamefont{Gottesman}} \bibnamefont{and}
  \bibinfo{author}{\bibfnamefont{I.~L.} \bibnamefont{Chuang}},
  \bibinfo{journal}{Nature} \textbf{\bibinfo{volume}{402}},
  \bibinfo{pages}{390} (\bibinfo{year}{1999}).

\bibitem[{\citenamefont{Cirac et~al.}(1997)\citenamefont{Cirac, Zoller, Kimble,
  and Mabuchi}}]{storage}
\bibinfo{author}{\bibfnamefont{J.~I.} \bibnamefont{Cirac}},
  \bibinfo{author}{\bibfnamefont{P.}~\bibnamefont{Zoller}},
  \bibinfo{author}{\bibfnamefont{H.~J.} \bibnamefont{Kimble}},
  \bibnamefont{and} \bibinfo{author}{\bibfnamefont{H.}~\bibnamefont{Mabuchi}},
  \bibinfo{journal}{Phys. Rev. Lett.} \textbf{\bibinfo{volume}{78}},
  \bibinfo{pages}{3221} (\bibinfo{year}{1997}).

\bibitem[{\citenamefont{Duan et~al.}(2001)\citenamefont{Duan, Lukin, Cirac, and
  Zoller}}]{derid}
\bibinfo{author}{\bibfnamefont{L.~M.} \bibnamefont{Duan}},
  \bibinfo{author}{\bibfnamefont{M.~D.} \bibnamefont{Lukin}},
  \bibinfo{author}{\bibfnamefont{J.~I.} \bibnamefont{Cirac}}, \bibnamefont{and}
  \bibinfo{author}{\bibfnamefont{P.}~\bibnamefont{Zoller}},
  \bibinfo{journal}{Nature} \textbf{\bibinfo{volume}{414}},
  \bibinfo{pages}{413} (\bibinfo{year}{2001}).

\bibitem[{\citenamefont{Boschi et~al.}(1998)\citenamefont{Boschi, Branca,
  De~Martini, Hardy, and Popescu}}]{boschi}
\bibinfo{author}{\bibfnamefont{D.}~\bibnamefont{Boschi}},
  \bibinfo{author}{\bibfnamefont{S.}~\bibnamefont{Branca}},
  \bibinfo{author}{\bibfnamefont{F.}~\bibnamefont{De~Martini}},
  \bibinfo{author}{\bibfnamefont{L.}~\bibnamefont{Hardy}}, \bibnamefont{and}
  \bibinfo{author}{\bibfnamefont{S.}~\bibnamefont{Popescu}},
  \bibinfo{journal}{Phys. Rev. Lett.} \textbf{\bibinfo{volume}{80}},
  \bibinfo{pages}{1121} (\bibinfo{year}{1998}).

\bibitem[{\citenamefont{Bouwmeester et~al.}(1997)\citenamefont{Bouwmeester,
  Pan, Mattle, Eibl, Weinfurter, and Zeilinger}}]{telezei}
\bibinfo{author}{\bibfnamefont{D.}~\bibnamefont{Bouwmeester}},
  \bibinfo{author}{\bibfnamefont{J.-W.} \bibnamefont{Pan}},
  \bibinfo{author}{\bibfnamefont{K.}~\bibnamefont{Mattle}},
  \bibinfo{author}{\bibfnamefont{M.}~\bibnamefont{Eibl}},
  \bibinfo{author}{\bibfnamefont{H.}~\bibnamefont{Weinfurter}},
  \bibnamefont{and}
  \bibinfo{author}{\bibfnamefont{A.}~\bibnamefont{Zeilinger}},
  \bibinfo{journal}{Nature} \textbf{\bibinfo{volume}{390}},
  \bibinfo{pages}{575} (\bibinfo{year}{1997}).

\bibitem[{\citenamefont{Furusawa et~al.}(1998)\citenamefont{Furusawa,
  S{\o}rensen, Braunstein, Fuchs, Kimble, and Polzik}}]{furuscience}
\bibinfo{author}{\bibfnamefont{A.}~\bibnamefont{Furusawa}},
  \bibinfo{author}{\bibfnamefont{J.~L.} \bibnamefont{S{\o}rensen}},
  \bibinfo{author}{\bibfnamefont{S.~L.} \bibnamefont{Braunstein}},
  \bibinfo{author}{\bibfnamefont{C.~A.} \bibnamefont{Fuchs}},
  \bibinfo{author}{\bibfnamefont{H.~J.} \bibnamefont{Kimble}},
  \bibnamefont{and} \bibinfo{author}{\bibfnamefont{E.~S.}
  \bibnamefont{Polzik}}, \bibinfo{journal}{Science}
  \textbf{\bibinfo{volume}{282}}, \bibinfo{pages}{706} (\bibinfo{year}{1998}).

\bibitem[{\citenamefont{{M. Riebe {\it et al.}}}(2004)}]{teleatom1}
\bibinfo{author}{\bibnamefont{{M. Riebe {\it et al.}}}},
  \bibinfo{journal}{Nature} \textbf{\bibinfo{volume}{429}},
  \bibinfo{pages}{734} (\bibinfo{year}{2004}).

\bibitem[{\citenamefont{{M. D. Barrett {\it et al.}}}(2004)}]{teleatom2}
\bibinfo{author}{\bibnamefont{{M. D. Barrett {\it et al.}}}},
  \bibinfo{journal}{Nature} \textbf{\bibinfo{volume}{429}},
  \bibinfo{pages}{737} (\bibinfo{year}{2004}).

\bibitem[{\citenamefont{Yonezawa et~al.}(2004)\citenamefont{Yonezawa, Aoki, and
  Furusawa}}]{naturusawa}
\bibinfo{author}{\bibfnamefont{H.}~\bibnamefont{Yonezawa}},
  \bibinfo{author}{\bibfnamefont{T.}~\bibnamefont{Aoki}}, \bibnamefont{and}
  \bibinfo{author}{\bibfnamefont{A.}~\bibnamefont{Furusawa}},
  \bibinfo{journal}{Nature} \textbf{\bibinfo{volume}{431}},
  \bibinfo{pages}{430} (\bibinfo{year}{2004}).

\bibitem[{\citenamefont{{N. Takei {\it et al.}}}(2005)}]{telesqz1}
\bibinfo{author}{\bibnamefont{{N. Takei {\it et al.}}}},
  \bibinfo{journal}{Phys. Rev. A} \textbf{\bibinfo{volume}{72}},
  \bibinfo{pages}{042304} (\bibinfo{year}{2005}).

\bibitem[{\citenamefont{Yonezawa et~al.}(2007)\citenamefont{Yonezawa,
  Braunstein, and Furusawa}}]{telesqz2}
\bibinfo{author}{\bibfnamefont{H.}~\bibnamefont{Yonezawa}},
  \bibinfo{author}{\bibfnamefont{S.~L.} \bibnamefont{Braunstein}},
  \bibnamefont{and} \bibinfo{author}{\bibfnamefont{A.}~\bibnamefont{Furusawa}},
  \bibinfo{journal}{Phys. Rev. Lett.} \textbf{\bibinfo{volume}{99}},
  \bibinfo{pages}{110503} (\bibinfo{year}{2007}).

\bibitem[{\citenamefont{{K. Honda {\it et al.}}}(2008)}]{memorysqz1}
\bibinfo{author}{\bibnamefont{{K. Honda {\it et al.}}}},
  \bibinfo{journal}{Phys. Rev. Lett.} \textbf{\bibinfo{volume}{100}},
  \bibinfo{pages}{093601} (\bibinfo{year}{2008}).

\bibitem[{\citenamefont{Appel et~al.}(2008)\citenamefont{Appel, Figueroa,
  Korystov, Lobino, and Lvovsky}}]{memorysqz2}
\bibinfo{author}{\bibfnamefont{J.}~\bibnamefont{Appel}},
  \bibinfo{author}{\bibfnamefont{E.}~\bibnamefont{Figueroa}},
  \bibinfo{author}{\bibfnamefont{D.}~\bibnamefont{Korystov}},
  \bibinfo{author}{\bibfnamefont{M.}~\bibnamefont{Lobino}}, \bibnamefont{and}
  \bibinfo{author}{\bibfnamefont{A.~I.} \bibnamefont{Lvovsky}},
  \bibinfo{journal}{Phys. Rev. Lett.} \textbf{\bibinfo{volume}{100}},
  \bibinfo{pages}{093602} (\bibinfo{year}{2008}).

\bibitem[{\citenamefont{Choi et~al.}(2008)\citenamefont{Choi, Deng, Laurat, and
  Kimble}}]{memolaurat}
\bibinfo{author}{\bibfnamefont{K.}~\bibnamefont{Choi}},
  \bibinfo{author}{\bibfnamefont{H.}~\bibnamefont{Deng}},
  \bibinfo{author}{\bibfnamefont{J.}~\bibnamefont{Laurat}}, \bibnamefont{and}
  \bibinfo{author}{\bibfnamefont{H.~J.} \bibnamefont{Kimble}},
  \bibinfo{journal}{Nature} \textbf{\bibinfo{volume}{452}}, \bibinfo{pages}{67}
  (\bibinfo{year}{2008}).

\bibitem[{\citenamefont{{T. Chaneli\`ere {\it et al.}}}(2005)}]{remotememory}
\bibinfo{author}{\bibnamefont{{T. Chaneli\`ere {\it et al.}}}},
  \bibinfo{journal}{Nature} \textbf{\bibinfo{volume}{438}},
  \bibinfo{pages}{833} (\bibinfo{year}{2005}).

\bibitem[{\citenamefont{Olmschenk et~al.}(2009)\citenamefont{Olmschenk,
  Matsukevich, Maunz, Hayes, Duan, and Monroe}}]{memsaab}
\bibinfo{author}{\bibfnamefont{S.}~\bibnamefont{Olmschenk}},
  \bibinfo{author}{\bibfnamefont{D.~N.} \bibnamefont{Matsukevich}},
  \bibinfo{author}{\bibfnamefont{P.}~\bibnamefont{Maunz}},
  \bibinfo{author}{\bibfnamefont{D.}~\bibnamefont{Hayes}},
  \bibinfo{author}{\bibfnamefont{L.-M.} \bibnamefont{Duan}}, \bibnamefont{and}
  \bibinfo{author}{\bibfnamefont{C.}~\bibnamefont{Monroe}},
  \bibinfo{journal}{Science} \textbf{\bibinfo{volume}{323}},
  \bibinfo{pages}{486} (\bibinfo{year}{2009}).

\bibitem[{\citenamefont{Hedges et~al.}(2010)\citenamefont{Hedges, Longdell, Li,
  and Sellars}}]{memory10}
\bibinfo{author}{\bibfnamefont{M.~P.} \bibnamefont{Hedges}},
  \bibinfo{author}{\bibfnamefont{J.~J.} \bibnamefont{Longdell}},
  \bibinfo{author}{\bibfnamefont{Y.}~\bibnamefont{Li}}, \bibnamefont{and}
  \bibinfo{author}{\bibfnamefont{M.~J.} \bibnamefont{Sellars}},
  \bibinfo{journal}{Nature} \textbf{\bibinfo{volume}{465}},
  \bibinfo{pages}{1052} (\bibinfo{year}{2010}).

\bibitem[{\citenamefont{Julsgaard et~al.}(2004)\citenamefont{Julsgaard,
  Sherson, Cirac, Fiur\'{a}\u{s}ek, and Polzik}}]{memorypolzik}
\bibinfo{author}{\bibfnamefont{B.}~\bibnamefont{Julsgaard}},
  \bibinfo{author}{\bibfnamefont{J.}~\bibnamefont{Sherson}},
  \bibinfo{author}{\bibfnamefont{J.~I.} \bibnamefont{Cirac}},
  \bibinfo{author}{\bibfnamefont{J.}~\bibnamefont{Fiur\'{a}\u{s}ek}},
  \bibnamefont{and} \bibinfo{author}{\bibfnamefont{E.~S.}
  \bibnamefont{Polzik}}, \bibinfo{journal}{Nature}
  \textbf{\bibinfo{volume}{432}}, \bibinfo{pages}{482} (\bibinfo{year}{2004}).

\bibitem[{\citenamefont{Sherson et~al.}(2006)\citenamefont{Sherson, Krauter,
  Olsson, Julsgaard, Hammerer, Cirac, and Polzik}}]{telepolzik}
\bibinfo{author}{\bibfnamefont{J.~F.} \bibnamefont{Sherson}},
  \bibinfo{author}{\bibfnamefont{H.}~\bibnamefont{Krauter}},
  \bibinfo{author}{\bibfnamefont{R.~K.} \bibnamefont{Olsson}},
  \bibinfo{author}{\bibfnamefont{B.}~\bibnamefont{Julsgaard}},
  \bibinfo{author}{\bibfnamefont{K.}~\bibnamefont{Hammerer}},
  \bibinfo{author}{\bibfnamefont{J.~I.} \bibnamefont{Cirac}}, \bibnamefont{and}
  \bibinfo{author}{\bibfnamefont{E.~S.} \bibnamefont{Polzik}},
  \bibinfo{journal}{Nature} \textbf{\bibinfo{volume}{443}},
  \bibinfo{pages}{557} (\bibinfo{year}{2006}).

\bibitem[{\citenamefont{{K. Jensen {\it et al.}}}(2011)}]{fernnp}
\bibinfo{author}{\bibnamefont{{K. Jensen {\it et al.}}}},
  \bibinfo{journal}{Nature Phys.} \textbf{\bibinfo{volume}{7}},
  \bibinfo{pages}{13} (\bibinfo{year}{2011}).

\bibitem[{\citenamefont{Krauter et~al.}(2013)\citenamefont{Krauter, Salart,
  Muschik, Petersen, Shen, Fernholz, and Polzik}}]{fernnp13}
\bibinfo{author}{\bibfnamefont{H.}~\bibnamefont{Krauter}},
  \bibinfo{author}{\bibfnamefont{D.}~\bibnamefont{Salart}},
  \bibinfo{author}{\bibfnamefont{C.~A.} \bibnamefont{Muschik}},
  \bibinfo{author}{\bibfnamefont{J.~M.} \bibnamefont{Petersen}},
  \bibinfo{author}{\bibfnamefont{H.}~\bibnamefont{Shen}},
  \bibinfo{author}{\bibfnamefont{T.}~\bibnamefont{Fernholz}}, \bibnamefont{and}
  \bibinfo{author}{\bibfnamefont{E.~S.} \bibnamefont{Polzik}},
  \bibinfo{journal}{Nature Phys.} \textbf{\bibinfo{volume}{9}},
  \bibinfo{pages}{400} (\bibinfo{year}{2013}).

\bibitem[{\citenamefont{Lee et~al.}(2011)\citenamefont{Lee, Benichi, Takeno,
  Takeda, Webb, Huntington, and Furusawa}}]{furuscience12}
\bibinfo{author}{\bibfnamefont{N.}~\bibnamefont{Lee}},
  \bibinfo{author}{\bibfnamefont{H.}~\bibnamefont{Benichi}},
  \bibinfo{author}{\bibfnamefont{Y.}~\bibnamefont{Takeno}},
  \bibinfo{author}{\bibfnamefont{S.}~\bibnamefont{Takeda}},
  \bibinfo{author}{\bibfnamefont{J.}~\bibnamefont{Webb}},
  \bibinfo{author}{\bibfnamefont{E.}~\bibnamefont{Huntington}},
  \bibnamefont{and} \bibinfo{author}{\bibfnamefont{A.}~\bibnamefont{Furusawa}},
  \bibinfo{journal}{Science} \textbf{\bibinfo{volume}{332}},
  \bibinfo{pages}{330} (\bibinfo{year}{2011}).

\bibitem[{\citenamefont{{X.-S. Ma {\it et al.}}}(2012)}]{zeil144}
\bibinfo{author}{\bibnamefont{{X.-S. Ma {\it et al.}}}},
  \bibinfo{journal}{Nature} \textbf{\bibinfo{volume}{489}},
  \bibinfo{pages}{269} (\bibinfo{year}{2012}).

\bibitem[{\citenamefont{Braunstein and van Loock}(2005)}]{brareview}
\bibinfo{author}{\bibfnamefont{S.~L.} \bibnamefont{Braunstein}}
  \bibnamefont{and} \bibinfo{author}{\bibfnamefont{P.}~\bibnamefont{van
  Loock}}, \bibinfo{journal}{Rev. Mod. Phys.} \textbf{\bibinfo{volume}{77}},
  \bibinfo{pages}{513} (\bibinfo{year}{2005}).

\bibitem[{\citenamefont{Cerf et~al.}(2007)\citenamefont{Cerf, Leuchs, and
  Polzik}}]{book}
\bibinfo{editor}{\bibfnamefont{N.}~\bibnamefont{Cerf}},
  \bibinfo{editor}{\bibfnamefont{G.}~\bibnamefont{Leuchs}}, \bibnamefont{and}
  \bibinfo{editor}{\bibfnamefont{E.~S.} \bibnamefont{Polzik}}, eds.,
  \emph{\bibinfo{title}{Quantum Information with Continuous Variables of Atoms
  and Light}} (\bibinfo{publisher}{Imperial College Press, London},
  \bibinfo{year}{2007}).

\bibitem[{\citenamefont{Furusawa and Takei}(2007)}]{cubic}
\bibinfo{author}{\bibfnamefont{A.}~\bibnamefont{Furusawa}} \bibnamefont{and}
  \bibinfo{author}{\bibfnamefont{N.}~\bibnamefont{Takei}},
  \bibinfo{journal}{Phys. Rep.} \textbf{\bibinfo{volume}{443}},
  \bibinfo{pages}{97} (\bibinfo{year}{2007}).

\bibitem[{\citenamefont{Hammerer et~al.}(2010)\citenamefont{Hammerer,
  S\o{}rensen, and Polzik}}]{hamrmp}
\bibinfo{author}{\bibfnamefont{K.}~\bibnamefont{Hammerer}},
  \bibinfo{author}{\bibfnamefont{A.~S.} \bibnamefont{S\o{}rensen}},
  \bibnamefont{and} \bibinfo{author}{\bibfnamefont{E.~S.}
  \bibnamefont{Polzik}}, \bibinfo{journal}{Rev. Mod. Phys.}
  \textbf{\bibinfo{volume}{82}}, \bibinfo{pages}{1041} (\bibinfo{year}{2010}).

\bibitem[{\citenamefont{Kimble}(2008)}]{qinternet}
\bibinfo{author}{\bibfnamefont{H.~J.} \bibnamefont{Kimble}},
  \bibinfo{journal}{Nature} \textbf{\bibinfo{volume}{453}},
  \bibinfo{pages}{1023} (\bibinfo{year}{2008}).

\bibitem[{\citenamefont{Uhlmann}(1976)}]{uhlmann}
\bibinfo{author}{\bibfnamefont{A.}~\bibnamefont{Uhlmann}},
  \bibinfo{journal}{Rep. Math. Phys.} \textbf{\bibinfo{volume}{9}},
  \bibinfo{pages}{273} (\bibinfo{year}{1976}).

\bibitem[{\citenamefont{Jozsa}(1994)}]{jozsa}
\bibinfo{author}{\bibfnamefont{R.}~\bibnamefont{Jozsa}}, \bibinfo{journal}{J.
  Mod. Opt.} \textbf{\bibinfo{volume}{41}}, \bibinfo{pages}{2315}
  (\bibinfo{year}{1994}).

\bibitem[{\citenamefont{Braunstein et~al.}(2000)\citenamefont{Braunstein,
  Fuchs, and Kimble}}]{brajmo}
\bibinfo{author}{\bibfnamefont{S.~L.} \bibnamefont{Braunstein}},
  \bibinfo{author}{\bibfnamefont{C.~A.} \bibnamefont{Fuchs}}, \bibnamefont{and}
  \bibinfo{author}{\bibfnamefont{H.~J.} \bibnamefont{Kimble}},
  \bibinfo{journal}{J. Mod. Opt.} \textbf{\bibinfo{volume}{47}},
  \bibinfo{pages}{267} (\bibinfo{year}{2000}).

\bibitem[{\citenamefont{Popescu}(1994)}]{popqub}
\bibinfo{author}{\bibfnamefont{S.}~\bibnamefont{Popescu}},
  \bibinfo{journal}{Phys. Rev. Lett.} \textbf{\bibinfo{volume}{72}},
  \bibinfo{pages}{797} (\bibinfo{year}{1994}).

\bibitem[{\citenamefont{Gisin}(1996)}]{gispla}
\bibinfo{author}{\bibfnamefont{N.}~\bibnamefont{Gisin}},
  \bibinfo{journal}{Phys. Lett. A} \textbf{\bibinfo{volume}{210}},
  \bibinfo{pages}{157} (\bibinfo{year}{1996}).

\bibitem[{\citenamefont{Horodecki et~al.}(1996)\citenamefont{Horodecki,
  Horodecki, and Horodecki}}]{horobell}
\bibinfo{author}{\bibfnamefont{R.}~\bibnamefont{Horodecki}},
  \bibinfo{author}{\bibfnamefont{M.}~\bibnamefont{Horodecki}},
  \bibnamefont{and}
  \bibinfo{author}{\bibfnamefont{P.}~\bibnamefont{Horodecki}},
  \bibinfo{journal}{Phys. Lett. A} \textbf{\bibinfo{volume}{222}},
  \bibinfo{pages}{21} (\bibinfo{year}{1996}).

\bibitem[{\citenamefont{Chiribella and Xie}(2013)}]{giulionew}
\bibinfo{author}{\bibfnamefont{G.}~\bibnamefont{Chiribella}} \bibnamefont{and}
  \bibinfo{author}{\bibfnamefont{J.}~\bibnamefont{Xie}},
  \bibinfo{journal}{Phys. Rev. Lett.} \textbf{\bibinfo{volume}{110}},
  \bibinfo{pages}{213602} (\bibinfo{year}{2013}).

\bibitem[{\citenamefont{Ho et~al.}(2013)\citenamefont{Ho, Bancal, and
  Scarani}}]{bancalnew}
\bibinfo{author}{\bibfnamefont{M.}~\bibnamefont{Ho}},
  \bibinfo{author}{\bibfnamefont{J.-D.} \bibnamefont{Bancal}},
  \bibnamefont{and} \bibinfo{author}{\bibfnamefont{V.}~\bibnamefont{Scarani}},
  \bibinfo{journal}{arXiv:1308.0084}  (\bibinfo{year}{2013}).

\bibitem[{\citenamefont{Bell}(1964)}]{bell}
\bibinfo{author}{\bibfnamefont{J.~S.} \bibnamefont{Bell}},
  \bibinfo{journal}{Physics} \textbf{\bibinfo{volume}{1}}, \bibinfo{pages}{195}
  (\bibinfo{year}{1964}).

\bibitem[{\citenamefont{Brunner et~al.}(2013)\citenamefont{Brunner, Cavalcanti,
  Pironio, Scarani, and Wehner}}]{bellrev}
\bibinfo{author}{\bibfnamefont{N.}~\bibnamefont{Brunner}},
  \bibinfo{author}{\bibfnamefont{D.}~\bibnamefont{Cavalcanti}},
  \bibinfo{author}{\bibfnamefont{S.}~\bibnamefont{Pironio}},
  \bibinfo{author}{\bibfnamefont{V.}~\bibnamefont{Scarani}}, \bibnamefont{and}
  \bibinfo{author}{\bibfnamefont{S.}~\bibnamefont{Wehner}},
  \bibinfo{journal}{arXiv:1303.2849}  (\bibinfo{year}{2013}).

\bibitem[{\citenamefont{Massar and Popescu}(1995)}]{benchqub}
\bibinfo{author}{\bibfnamefont{S.}~\bibnamefont{Massar}} \bibnamefont{and}
  \bibinfo{author}{\bibfnamefont{S.}~\bibnamefont{Popescu}},
  \bibinfo{journal}{Phys. Rev. Lett.} \textbf{\bibinfo{volume}{74}},
  \bibinfo{pages}{1259} (\bibinfo{year}{1995}).

\bibitem[{\citenamefont{Bruss and Macchiavello}(1999)}]{benchd}
\bibinfo{author}{\bibfnamefont{D.}~\bibnamefont{Bruss}} \bibnamefont{and}
  \bibinfo{author}{\bibfnamefont{C.}~\bibnamefont{Macchiavello}},
  \bibinfo{journal}{Phys. Lett. A} \textbf{\bibinfo{volume}{253}},
  \bibinfo{pages}{249} (\bibinfo{year}{1999}).

\bibitem[{\citenamefont{Hammerer et~al.}(2005)\citenamefont{Hammerer, Wolf,
  Polzik, and Cirac}}]{hammerer}
\bibinfo{author}{\bibfnamefont{K.}~\bibnamefont{Hammerer}},
  \bibinfo{author}{\bibfnamefont{M.~M.} \bibnamefont{Wolf}},
  \bibinfo{author}{\bibfnamefont{E.~S.} \bibnamefont{Polzik}},
  \bibnamefont{and} \bibinfo{author}{\bibfnamefont{J.~I.} \bibnamefont{Cirac}},
  \bibinfo{journal}{Phys. Rev. Lett.} \textbf{\bibinfo{volume}{94}},
  \bibinfo{pages}{150503} (\bibinfo{year}{2005}).

\bibitem[{\citenamefont{Adesso and Chiribella}(2008)}]{noi}
\bibinfo{author}{\bibfnamefont{G.}~\bibnamefont{Adesso}} \bibnamefont{and}
  \bibinfo{author}{\bibfnamefont{G.}~\bibnamefont{Chiribella}},
  \bibinfo{journal}{Phys. Rev. Lett.} \textbf{\bibinfo{volume}{100}},
  \bibinfo{pages}{170503} (\bibinfo{year}{2008}).

\bibitem[{\citenamefont{Owari et~al.}(2008)\citenamefont{Owari, Plenio, Polzik,
  Serafini, and Wolf}}]{owari}
\bibinfo{author}{\bibfnamefont{M.}~\bibnamefont{Owari}},
  \bibinfo{author}{\bibfnamefont{M.~B.} \bibnamefont{Plenio}},
  \bibinfo{author}{\bibfnamefont{E.~S.} \bibnamefont{Polzik}},
  \bibinfo{author}{\bibfnamefont{A.}~\bibnamefont{Serafini}}, \bibnamefont{and}
  \bibinfo{author}{\bibfnamefont{M.~M.} \bibnamefont{Wolf}},
  \bibinfo{journal}{New J. Phys.} \textbf{\bibinfo{volume}{10}},
  \bibinfo{pages}{113014} (\bibinfo{year}{2008}).

\bibitem[{\citenamefont{Calsamiglia et~al.}(2009)\citenamefont{Calsamiglia,
  Aspachs, Mu\~noz Tapia, and Bagan}}]{mariona}
\bibinfo{author}{\bibfnamefont{J.}~\bibnamefont{Calsamiglia}},
  \bibinfo{author}{\bibfnamefont{M.}~\bibnamefont{Aspachs}},
  \bibinfo{author}{\bibfnamefont{R.}~\bibnamefont{Mu\~noz Tapia}},
  \bibnamefont{and} \bibinfo{author}{\bibfnamefont{E.}~\bibnamefont{Bagan}},
  \bibinfo{journal}{Phys. Rev. A} \textbf{\bibinfo{volume}{79}},
  \bibinfo{pages}{050301(R)} (\bibinfo{year}{2009}).

\bibitem[{\citenamefont{Gu{\c{t}}\u{a}
  et~al.}(2010)\citenamefont{Gu{\c{t}}\u{a}, Bowles, and Adesso}}]{peterb}
\bibinfo{author}{\bibfnamefont{M.}~\bibnamefont{Gu{\c{t}}\u{a}}},
  \bibinfo{author}{\bibfnamefont{P.}~\bibnamefont{Bowles}}, \bibnamefont{and}
  \bibinfo{author}{\bibfnamefont{G.}~\bibnamefont{Adesso}},
  \bibinfo{journal}{Phys. Rev. A} \textbf{\bibinfo{volume}{82}},
  \bibinfo{pages}{042310} (\bibinfo{year}{2010}).

\bibitem[{\citenamefont{Weedbrook et~al.}(2012)\citenamefont{Weedbrook,
  Pirandola, Garcia-Patron, Cerf, Ralph, Shapiro, and Lloyd}}]{pirandolareview}
\bibinfo{author}{\bibfnamefont{C.}~\bibnamefont{Weedbrook}},
  \bibinfo{author}{\bibfnamefont{S.}~\bibnamefont{Pirandola}},
  \bibinfo{author}{\bibfnamefont{R.}~\bibnamefont{Garcia-Patron}},
  \bibinfo{author}{\bibfnamefont{N.~J.} \bibnamefont{Cerf}},
  \bibinfo{author}{\bibfnamefont{T.~C.} \bibnamefont{Ralph}},
  \bibinfo{author}{\bibfnamefont{J.~H.} \bibnamefont{Shapiro}},
  \bibnamefont{and} \bibinfo{author}{\bibfnamefont{S.}~\bibnamefont{Lloyd}},
  \bibinfo{journal}{Rev. Mod. Phys.} \textbf{\bibinfo{volume}{84}},
  \bibinfo{pages}{621} (\bibinfo{year}{2012}).

\bibitem[{\citenamefont{Adesso and Illuminati}(2007)}]{ourreview}
\bibinfo{author}{\bibfnamefont{G.}~\bibnamefont{Adesso}} \bibnamefont{and}
  \bibinfo{author}{\bibfnamefont{F.}~\bibnamefont{Illuminati}},
  \bibinfo{journal}{J. Phys. A: Math. Theor.} \textbf{\bibinfo{volume}{40}},
  \bibinfo{pages}{7821} (\bibinfo{year}{2007}).

\bibitem[{\citenamefont{{van Loock} and Braunstein}(2000)}]{network}
\bibinfo{author}{\bibfnamefont{P.}~\bibnamefont{{van Loock}}} \bibnamefont{and}
  \bibinfo{author}{\bibfnamefont{S.~L.} \bibnamefont{Braunstein}},
  \bibinfo{journal}{Phys. Rev. Lett.} \textbf{\bibinfo{volume}{84}},
  \bibinfo{pages}{3482} (\bibinfo{year}{2000}).

\bibitem[{\citenamefont{Adesso and Illuminati}(2005)}]{telepoppy}
\bibinfo{author}{\bibfnamefont{G.}~\bibnamefont{Adesso}} \bibnamefont{and}
  \bibinfo{author}{\bibfnamefont{F.}~\bibnamefont{Illuminati}},
  \bibinfo{journal}{Phys. Rev. Lett.} \textbf{\bibinfo{volume}{95}},
  \bibinfo{pages}{150503} (\bibinfo{year}{2005}).

\bibitem[{\citenamefont{Zhang et~al.}(2009)\citenamefont{Zhang, Adesso, Xie,
  and Peng}}]{teamwork}
\bibinfo{author}{\bibfnamefont{J.}~\bibnamefont{Zhang}},
  \bibinfo{author}{\bibfnamefont{G.}~\bibnamefont{Adesso}},
  \bibinfo{author}{\bibfnamefont{C.}~\bibnamefont{Xie}}, \bibnamefont{and}
  \bibinfo{author}{\bibfnamefont{K.}~\bibnamefont{Peng}},
  \bibinfo{journal}{Phys. Rev. Lett.} \textbf{\bibinfo{volume}{103}},
  \bibinfo{pages}{070501} (\bibinfo{year}{2009}).

\bibitem[{noi()}]{noinote}
\bibinfo{note}{In Ref.~\cite{noi}, the state reprepared by Bob was restricted
  to be a pure squeezed state belonging to the same set as the input state. If
  one optimizes the CFT of \cite{noi} by allowing for an arbitrary
  repreparation, one finds that the optimal state Bob can prepare is
  non-Gaussian, and the corresponding deterministic benchmark for teleporting
  undisplaced, unrotated squeezed vacuum states with totally random $s$ turns
  out to be $\approx 0.832$.}

\bibitem[{hay()}]{hayashi}
\bibinfo{note}{M. Hayashi, talk given at \emph{9th International Conference on
  Squeezed States and Uncertainty Relations, ICSSUR} (2005).}

\bibitem[{epa()}]{epaps}
\bibinfo{note}{See the Supplemental Material at [EPAPS] for technical
  derivations.}

\bibitem[{\citenamefont{Hausladen and Wootters}(1994)}]{square}
\bibinfo{author}{\bibfnamefont{P.}~\bibnamefont{Hausladen}} \bibnamefont{and}
  \bibinfo{author}{\bibfnamefont{W.~K.} \bibnamefont{Wootters}},
  \bibinfo{journal}{Journal of Modern Optics} \textbf{\bibinfo{volume}{41}},
  \bibinfo{pages}{2385} (\bibinfo{year}{1994}).

\bibitem[{\citenamefont{Hausladen et~al.}(1996)\citenamefont{Hausladen, Jozsa,
  Schumacher, Westmoreland, and Wootters}}]{srm}
\bibinfo{author}{\bibfnamefont{P.}~\bibnamefont{Hausladen}},
  \bibinfo{author}{\bibfnamefont{R.}~\bibnamefont{Jozsa}},
  \bibinfo{author}{\bibfnamefont{B.}~\bibnamefont{Schumacher}},
  \bibinfo{author}{\bibfnamefont{M.}~\bibnamefont{Westmoreland}},
  \bibnamefont{and} \bibinfo{author}{\bibfnamefont{W.~K.}
  \bibnamefont{Wootters}}, \bibinfo{journal}{Phys. Rev. A}
  \textbf{\bibinfo{volume}{54}}, \bibinfo{pages}{1869} (\bibinfo{year}{1996}).

\bibitem[{\citenamefont{Abramowitz and Stegun}(1964)}]{borntobe}
\bibinfo{author}{\bibfnamefont{M.}~\bibnamefont{Abramowitz}} \bibnamefont{and}
  \bibinfo{author}{\bibfnamefont{I.}~\bibnamefont{Stegun}},
  \emph{\bibinfo{title}{Handbook of Mathematical Functions}}
  (\bibinfo{publisher}{Dover}, \bibinfo{address}{New York},
  \bibinfo{year}{1964}).

\bibitem[{\citenamefont{Gilmore}(1972)}]{gilmore}
\bibinfo{author}{\bibfnamefont{R.}~\bibnamefont{Gilmore}},
  \bibinfo{journal}{Annals of Physics} \textbf{\bibinfo{volume}{74}},
  \bibinfo{pages}{391} (\bibinfo{year}{1972}).

\bibitem[{\citenamefont{Perelomov}(1972)}]{perelomov}
\bibinfo{author}{\bibfnamefont{A.~M.} \bibnamefont{Perelomov}},
  \bibinfo{journal}{Comm. Math. Phys.} \textbf{\bibinfo{volume}{26}},
  \bibinfo{pages}{222} (\bibinfo{year}{1972}).

\bibitem[{\citenamefont{Perelomov}(1986)}]{perelomovbook}
\bibinfo{author}{\bibfnamefont{A.}~\bibnamefont{Perelomov}}, \emph{\bibinfo{title}{Generalized coherent states and their applications}}
  (\bibinfo{publisher}{Springer}, \bibinfo{year}{1986}).

\bibitem[{\citenamefont{Ali et~al.}(2000)\citenamefont{Ali, Antoine, and
  Gazeau}}]{ali}
\bibinfo{author}{\bibfnamefont{S.~T.} \bibnamefont{Ali}},
  \bibinfo{author}{\bibfnamefont{J.-P.} \bibnamefont{Antoine}},
  \bibnamefont{and} \bibinfo{author}{\bibfnamefont{J.~P.}
  \bibnamefont{Gazeau}}, \emph{\bibinfo{title}{Coherent states, wavelets and
  their generalizations}} (\bibinfo{publisher}{Springer},
  \bibinfo{year}{2000}).

\bibitem[{\citenamefont{Berceanu}(2009)}]{berceanu}
\bibinfo{author}{\bibfnamefont{S.}~\bibnamefont{Berceanu}},
  \bibinfo{journal}{AIP Conf. Proc.} \textbf{\bibinfo{volume}{1079}},
  \bibinfo{pages}{67} (\bibinfo{year}{2009}).

\bibitem[{\citenamefont{{van Loock}}(2002)}]{vloko}
\bibinfo{author}{\bibfnamefont{P.}~\bibnamefont{{van Loock}}},
  \bibinfo{journal}{Fortschr. Phys.} \textbf{\bibinfo{volume}{50}},
  \bibinfo{pages}{1177} (\bibinfo{year}{2002}).

\bibitem[{\citenamefont{{T. Eberle {\it et al.}}}(2010)}]{schnabel}
\bibinfo{author}{\bibnamefont{{T. Eberle {\it et al.}}}},
  \bibinfo{journal}{Phys. Rev. Lett.} \textbf{\bibinfo{volume}{104}},
  \bibinfo{pages}{251102} (\bibinfo{year}{2010}).

\bibitem[{\citenamefont{Eberle et~al.}(2013)\citenamefont{Eberle, H\"{a}ndchen,
  and Schnabel}}]{schnabel13}
\bibinfo{author}{\bibfnamefont{T.}~\bibnamefont{Eberle}},
  \bibinfo{author}{\bibfnamefont{V.}~\bibnamefont{H\"{a}ndchen}},
  \bibnamefont{and} \bibinfo{author}{\bibfnamefont{R.}~\bibnamefont{Schnabel}},
  \bibinfo{journal}{Opt. Express} \textbf{\bibinfo{volume}{21}},
  \bibinfo{pages}{11546} (\bibinfo{year}{2013}).

\end{thebibliography}

\begin{thebibliography}{30}
\bibitem[S1]{Sgiulionew}  G. Chiribella and J. Xie, Phys. Rev. Lett. {\bf 110}, 213602 (2013).
\bibitem[S2]{Sklee}   V. Klee, Canad. J. Math. {\bf 16}, 517 (1963).
\bibitem[S3]{Ssquare} P. Hausladen and W. K. Wootters,  J. Mod. Opt. {41},  2385-2390 (1994).
\bibitem[S4]{Ssrm} P. Hausladen, R. Jozsa, B. Schumacher, M. Westmoreland, and W. K. Wootters, Phys. Rev. A {\bf 54}, 1869 (1996).
\bibitem[S5]{Sperelomov2} A. M. Perelomov, \emph{Generalized coherent states and their applications} (Springer-Verlag, Berlin, 1986).
\bibitem[S6]{Sali}  S. T. Ali, J. P. Antoine and J. P. Gazeau, \emph{Coherent States, Wavelets and Their Generalizations} (Springer-Verlag, New York, 2000).
\bibitem[S7]{Sgilmore} R. Gilmore, Ann. of Phys. {\bf 74} 391 (1972).
\bibitem[S8]{Sperelomov} A. M. Perelomov, Comm. Math. Phys. {\bf 26}, 222 (1972).
\bibitem[S9]{Sberceanu} S. Berceanu, AIP Conf. Proc. {\bf 1079}, 67 (2009).
\bibitem[S10]{Snortonetal}P. Kral, J. Mod. Opt. {\bf 37}, 889 (1990); C. M. A. Dantas, N. G. de Almeida, and B. Baseia, Braz. J. Phys. {\bf 28}, 462 (1998).
\bibitem[S11]{Shelstrom} C. W. Helstrom, \emph{Quantum detection and estimation theory} (Academic Press, New York, 1976).
\bibitem[S12]{Sholevo} A. S. Holevo, \emph{Probabilistic and Statistical Aspects of Quantum
Theory} (North Holland, Amsterdam, 1982).
\bibitem[S13]{Sozawa}  M. Ozawa, in \emph{Research Reports on Information Sciences, Series
A: Mathematical Sciences, N. 74}, Department of Information
Sciences, Tokyo Institute of Technology (1980).
\bibitem[S14]{Shayashi} M. Hayashi, talk given at \emph{9th International Conference
on Squeezed States and Uncertainty Relations, ICSSUR}  (2005).
\end{thebibliography}


\clearpage
\begin{widetext}

\subsection*{{\large Supplemental Material: Quantum benchmarks for pure single-mode Gaussian states}}

\appendix

\appendix

\section{Proof of Eq. (\element{3a}): benchmark for squeezed vacuum states}\label{sec:squeez}
Let us expand the squeezed  states as
\begin{align*}
|\xi \>  &:=     e^{\frac12(\xi \hat a^{\dag  2}  -  \xi^\ast   \hat a^2)}  |0\>  \\
&= \frac 1 {\sqrt {\cosh  s}}    \sum_{n=0}^\infty  \sqrt{   n -\frac 12 \choose n} \left({\tanh  s}  \right)^{n}  e^{in\theta}  |2 n\>\,, \qquad  \xi  =   s  e^{i\theta}
 \end{align*}
 using the notation $  {x \choose n}  :  =  \frac { x (x-1)  \dots (x-n+1)}{n!}$ for a generic $x \in\mathbb R$.
We now compute the average states of the ensembles $\left\{  |\xi\>  ,  p^S_\beta (\xi)\right\}$ and $\left\{|\xi\>\otimes |\xi\>, p^S_\beta  (\xi)  \right\}$, where $p^S_\beta (\xi)$ is the probability distribution
\begin{align*}
p^S_\beta(\xi)   =    \frac{  \beta     \sinh s ~    }{ (\cosh s)^{\beta + 1}} ~ \frac{ ~1}{2\pi}\,,\qquad \beta>  0,
\end{align*}
satisfying the normalization condition $\int \d s \d \theta ~ p_\beta^S (\xi) = 1$.
By explicit calculation, we find that the average states are
\begin{align}\label{uno}
\tau_\beta   : =  \int  \d s \d \theta ~p^S_\beta(\xi)   ~  |\xi\>\< \xi  |
    = \frac{\beta}{\beta + 1}  \sum_{  n=0}^\infty  \frac{  {   n - \frac 1 2  \choose n}  } {{    \frac{\beta + 1}2  + n  \choose n}}  ~  |2n\>\<  2n| ,
    \end{align}
    and
    \begin{align}\label{due}
\rho_\beta  :  =   &  \int \d s \d \theta ~ p^S_\beta(\xi)~    |  \xi\>\< \xi|  \otimes |\xi\>\<  \xi|    =  \frac{  \beta}{\beta + 2 }  \sum_{k=0}^{\infty}  \frac 1 {{  \frac {  \beta + 2}2  +  k  \choose k}}   | \Phi_k\>\<  \Phi_k|
  \end{align}
 where
 \begin{align}\label{ficappa}
 |\Phi_k\>
 &  : =  \sum_{n=0}^k  \sqrt{    {   k-n-\frac 12   \choose k-n}  {   n  - \frac 12  \choose n} } ~   |  2k-2n \>  |2n\>.
 \end{align}

Now, by Ref.~\cite{Sgiulionew} the probabilistic CFT is given by
\begin{equation*}
\fc^S (\beta)  =   \left\|  A_\beta  \right\|_\times\,, \qquad A_\beta : = \left(  I  \otimes \tau_\beta^{-1/2}\right)  \rho_\beta  \left( I\otimes \tau_\beta^{-1/2} \right)\, ,
\end{equation*}
where  $\|  A_{\beta } \|_{\times}$ is the {cross norm} of $A_\beta$,  defined as  $\left \| A_\beta\right\|_\times  :  =  \sup_{ \| \varphi\| =  \|\psi\|=1} \<  \varphi| \< \psi| A_\beta |\varphi\>|\psi\>$.  Using Eq. (\ref{due}), one can write the operator $A_\beta$ as
\begin{align*}
A_{\beta}  =    \frac{  \beta}{\beta + 2 }  \sum_{k=0}^{\infty}  \frac 1 {{  \frac {  \beta + 2}2  +  k  \choose k}}   ~  \left (I \otimes \tau_\beta^{-\frac 12}\right)  |\Phi_k\>\<  \Phi_k|   \left (I \otimes \tau_\beta^{-\frac 12}\right)
\end{align*}
where the vectors $  \left (I \otimes \tau_\beta^{-\frac 12}\right)  |\Phi_k\>$ are mutually orthogonal for different values of $k$, as can be seen by direct inspection using Eq.~(\ref{uno}).
Using this fact, one can compute  the eigenvalues of $A_\beta$, which are  given by
 \begin{align*}
 a_{\beta,k}  &=   \frac{\beta }{\beta+2}   \frac 1 {{  \frac {  \beta + 2}2  +  k  \choose k} }    \< \Phi_k|  (I \otimes \tau_\beta^{-1})   |\Phi_k\>    \\
 &  =   \frac{\beta+1 }{\beta+2}   \frac 1 {{  \frac {  \beta + 2}2  +  k  \choose k}}     \sum_{n=0}^k      {   k-n-\frac 1 2  \choose k-n}  {   \frac {\beta +1} 2  + n  \choose n}    \\
 &  =    \frac{\beta+1 }{\beta+2}   \frac {  (-1)^k  } {{  \frac {  \beta + 2}2  +  k  \choose k}}     \sum_{n=0}^k  { -\frac 1 2  \choose k-n}  {   -\frac {\beta +3} 2    \choose n}  \\
 &  =    \frac{\beta+1 }{\beta+2}~   \frac {  (-1)^k  } {{  \frac {  \beta + 2}2  +  k  \choose k}}         {   -\frac {\beta +4} 2    \choose k}   \\
   &  =  \frac{\beta+1 }{\beta+2}    ,
 \end{align*}
 (the fourth equality coming from the Chu-Vandermonde identity \cite{Sklee}). In other words, $A_{\beta}$  is proportional to a projector, with the proportionality constant $(\beta+  1)/(\beta+2)$.
 This proves that
 \begin{align*}
\left \| A_\beta\right\|_\times  & =  \sup_{ \| \varphi\| =  \|\psi\|=1} \<  \varphi| \< \psi| A_\beta |\varphi\>|\psi\>\\
   &   \le    \frac{\beta+1 }{\beta+2}  . \end{align*}
 On the other hand, observing that   $   \<  0| \< 0| A_\beta |0\>|0\> =   \frac{\beta+1 }{\beta+2}   $ we conclude that
 \begin{equation*}
 \fc^S (\beta)  = \left \| A_\beta\right\|_\times    = \frac {\beta +1}{\beta +2}.
 \end{equation*}

\section{The fidelity of the square-root measurement}

For the the ensemble of squeezed states $\{ |\xi\>  ,  p^S_\eta (\d^2 \xi) \}$ the square-root measurement is the POVM with operators  $  P_\eta  (\xi  )  =    p^S_\eta  (\xi)  \tau_\eta^{-\frac 12}  |  \xi\>\<  \xi|  \tau_\eta^{-\frac 12} $  \cite{Ssquare,Ssrm}.  Hence, the fidelity of the measure-and-prepare protocol based on measuring the square-root measurement and on re-preparing squeezed states is given by
\begin{align*}
\overline{F_{sr}}^S  &  =   \int \d s \d \theta   \int  \d s' \d \theta'  ~   p_\beta^S  (\xi)      ~ | \<  \xi|  \xi'\>|^2 ~  \<  \xi|  P_\eta  (\xi')  |\xi\>  \\
&  =       \int \d s \d \theta   \int  \d s' \d \theta'  ~   p_\beta^S  (\xi)        p_\beta^S  (\xi')   ~\Tr \left[     \left(  |\xi\>\<  \xi|^{\phantom{\frac 12}}_{\phantom{\eta}}  \otimes |\xi\>\< \xi|   \right)  \left(  |\xi'\>\<  \xi'|  \otimes   \tau_\eta^{-\frac 12}  |\xi'\>\<  \xi'|  \tau_\eta^{-\frac 12}  \right) \right]\\
&  =  \Tr  \left[ \rho_\beta  (  I  \otimes \tau_\eta^{-\frac 12})  \rho_\eta (  I  \otimes \tau_\eta^{-\frac 12})    \right]   \, .
\end{align*}
Inserting Eqs.~ (\ref{uno}),  (\ref{due}), and (\ref{ficappa}) into the last equation we then obtain
\begin{align*}
\overline{F_{sr}}^S    & =      \frac{\beta}{\beta +2}     \frac{\eta}{\eta +2}   \sum_{k=0}^{\infty} \frac 1 { {  \frac {  \beta + 2}2  +  k  \choose k}   {  \frac {  \eta + 2}2  +  k  \choose k}  }   \left|  \< \Phi_k|  \left (I \otimes \tau_\eta^{- \frac 12}  \right) |\Phi_k\>\right|^2 \\
    & =   \frac{\beta}{\beta +2}  \frac{ \eta+ 1}{\eta+2}    \sum_{k=0}^\infty  \frac 1 { {  \frac {  \beta + 2}2  +  k  \choose k}   {  \frac {  \eta + 2}2  +  k  \choose k}  }
      {\left|  \sum_{n=0}^k  {k-n -\frac 12 \choose  k-n}        \sqrt{ {   n-\frac 12\choose n  } {  \frac {\eta + 1} 2  +n \choose  n  } } \right|^2} \, ,
  \end{align*}
  which is the expression used in the main text.
 \section{Proof of Eq. (8):  benchmark for teleportation and storage of Gilmore-Perelomov coherent states}
Consider a generic Lie group $\grp G$, acting on a Hilbert space $\spc H$ through a unitary irreducible representation $\hat U:    g \mapsto \hat U_g$.   In the following we refer to the monographies \cite{Sperelomov2,Sali} for the background on coherent states and representation theory.    We will consider    \emph{Gilmore-Perelomov  coherent states}  \cite{Sgilmore,Sperelomov} and  $  |\varphi_g\> $ of the form $  |\varphi_g\> =   \hat U_g  |\varphi\>$, where  $|\varphi\>$ is a lowest weight vector for the representation $U$, i.e. a vector that is annihilated by all the negative roots of the Lie algebra      Examples of Gilmore-Perelomov coherent states  are the ordinary coherent states $|\alpha\>  =  \hat D(\alpha)|0\>$, associated to the Weyl-Heisenberg group, the squeezed states   $|\xi\>  =  \hat S (\xi) |0\>$, associated to the group $SU(1,1)$, and the displaced squeezed states $ |\alpha,\xi\> =  \hat D(\alpha) \hat S (\xi) |0\>$, associated to the Jacobi group.   Other examples, in finite dimensional quantum systems, are the pure states $ |\varphi_U\>  =  U  |\varphi_0\>$, $U\in SU(d)$ and the spin-coherent states $  |j,j ,\varphi, \vec n\>=    R^{(j)}_{\varphi,\vec n} |j,j\>$ associated to the rotation group $SO(3)$.

We now prove a general formula to compute the classical fidelity threshold for the teleportation and storage of Gilmore-Perelomov coherent states.    Among the possible measure-and-prepare strategies, we include probabilistic strategies based on abstention. To stress this fact, we refer to our  CFT as \emph{probabilistic CFT}.
We assume that the group is \emph{unimodular}---that is, it has a left- and right- invariant measure $\d g$---, and that the input state $|\varphi_g\>$ is given with prior probability
\begin{equation}\label{prob}
p_\gamma(\d g)   =  d_\gamma  |\<   \varphi_\gamma  |\varphi_{\gamma,g}\>   |^2  \d g     ,
\end{equation}
where  $ |\varphi_{\gamma,g}\>  :  = \hat U^\gamma_g |\varphi_\gamma\>\in\spc H_\gamma$  is a coherent state for some other irreducible representation $\hat U^\gamma :  g \to  \hat U^\gamma_g  $ and  $d_\gamma$ is a normalization constant, known as \emph{formal dimension}, given by
\begin{equation}\label{di}
d_\gamma :  =   \left(  \int   \d g ~     |\<   \varphi_\gamma        |\varphi_{\gamma,g}\>   |^2    \right)^{-1}  .
\end{equation}
Of course, in order for the probability distribution $p_\gamma (\d g)$ to be normalizable, the formal dimension $d_\gamma$ should not be zero  (i.e. the  integral in Eq. (\ref{di}) should not diverge).  Technically, irreducible representations with this property are called \emph{square-summable}.   Since we want $p_\gamma^S  (\d g)$ to be a probability distribution, in the following  we will always assume that the representation $\hat U_\gamma$ is square-summable.

In addition, we will always assume that that  the root system that makes $|\varphi_\gamma\>$ a lowest weight vector has been chosen to have the same structure constants of the root system that makes $|\varphi\>$ a lowest weight vector.  For example, this is what is done in quantum optics when one has two annihilation operators $\hat a$ and $\hat b$ for two different modes $A$ and $B$, that are chosen in such a way that $[\hat a, \hat a^\dag]  = I_A  $ and $[\hat b,\hat b^\dag]  =   I_B$.  With this choice, the negative roots of the Lie algebra representation associated to the product representations  $\hat U  \otimes \hat U_\gamma$  and  $\hat U  \otimes  \hat U\otimes  \hat U_\gamma$ will be the sum of the negative roots of the Lie algebra representation associated to $\hat U$ and $\hat U_\gamma$.   Hence,  the fact that both $|\varphi\>$ and $|\varphi_\gamma\>$ are annihilated by the negative roots, implies that also the states $|\varphi\>  |\varphi_\gamma\>$ and $|\varphi\>|\varphi\>  |\varphi_\gamma\>$ are annihilated by the negative roots, i.e. they are lowest weight vectors. In the example of quantum optics, this corresponds to the fact that  product of the vacuum states for two modes $a$ and $b$ is the vacuum state for the mode $a  +  b$.
Our choice of root system guarantees also that the product representations $ \hat U  \otimes \hat U_\gamma$ and $ \hat U  \otimes \hat U \otimes \hat U_\gamma$ are square summable: indeed, one has
\begin{align*}
d_{\hat U  \otimes \hat U \otimes \hat U_\gamma}  &  \equiv  \left(  \int   \d g ~         |\<   \varphi        |\varphi_g\>   |^4      |\<   \varphi_\gamma        |\varphi_{\gamma,g}\>   |^2    \right)^{-1}  \\
&  \ge   \left(  \int   \d g ~         |\<   \varphi        |\varphi_g\>   |^2      |\<   \varphi_\gamma        |\varphi_{\gamma,g}\>   |^2    \right)^{-1} \\
 &  \equiv  d_{\hat U  \otimes \hat U_\gamma}  \\
 &  \ge  \left(  \int   \d g ~             |\<   \varphi_\gamma        |\varphi_{\gamma,g}\>   |^2    \right)^{-1}\\
 &  \equiv  d_{\gamma}\\
  &  >  0 \, .
\end{align*}

With the above settings, we have the following result
\begin{Theo}[Benchmark for Gilmore-Perelomov coherent states]\label{theo:cft}
Let $\Lambda  =  \{ |\varphi_g\>,  p_\gamma (\d g) \}_{g\in\grp G}$ be an ensemble of Gilmore-Perelomov coherent states, with prior probability distribution of the form $p_\gamma(\d g)  = d_\gamma  |\<   \varphi_\gamma  |\varphi_{\gamma,g}\>   |^2  \d g $.
Then, the probabilistic CFT is
\begin{equation}\label{probcft}
\fc (\gamma)  =
\frac  {  \int \d g~      p_\gamma (g)    |\<   \varphi |   \varphi_g\> |^4  }  {  \int \d g~      p_\gamma (g)    |\<   \varphi |   \varphi_g\> |^2  }.
\end{equation}
\end{Theo}

\Proof By the result of Ref.~\cite{Sgiulionew}, the CFT is given by the cross norm
\begin{align*}
 \fc (\gamma) & =   \left\| A_\gamma     \right\|_\times   \qquad
 \left \{
 \begin{array}{ll} A_\gamma &: = (  I  \otimes \tau_\gamma^{-1/2})  \rho_\gamma ( I \otimes \tau_\gamma^{-1/2} )\\
   \tau_\gamma    &:=  \int  \d g ~  p_\gamma (  g) ~   |\varphi_g\>\<  \varphi_g|   \\
   \rho_\gamma  &  :  =  \int  \d g ~  p_\gamma (  g) ~   |\varphi_g\>\<  \varphi_g| \otimes  |\varphi_g\>\<  \varphi_g|
   \end{array}
   \right.  \\
   &\le  \|A_\gamma\|_{\infty}\\
   &  =  \min  \{\lambda  \ge 0 ~|~    \lambda  (I \otimes   \tau_\gamma)   \ge  \rho_\gamma \}
\end{align*}
Now, using Eq. (\ref{prob}),   we have
\begin{align}\label{botheqs}
\tau_\gamma  &  =  d_\gamma \Tr_{\gamma}    \left[  \left (   \int  \d g ~     |\varphi_g\>\<  \varphi_g| \otimes |\varphi_{\gamma,g}\>\< \varphi_{\gamma,g}|     \right)    (  I \otimes |\varphi_\gamma\>\<\varphi_\gamma|)\right ]\\
  \rho_\gamma  &  =  d_\gamma\Tr_{\gamma}\left[ \left(  \int  \d g ~   | \varphi_g\>\<  \varphi_g| \otimes  |\varphi_g\>\<  \varphi_g|  \otimes |\varphi_{\gamma,g}\>\< \varphi_{\gamma, g}|  \right)  \left(   I \otimes  I \otimes |\varphi_\gamma\>\<\varphi_\gamma| \right)    \right] \, ,
\end{align}
where $  \Tr_\gamma$ is a shorthand notation for the partial trace over the Hilbert space $\spc H_\gamma$.
Now,  since the state $|\varphi\>  |\varphi_\gamma\>$  (respectively, $|\varphi\>|\varphi\>  |\varphi_\gamma\>$) is a lowest weight vector,  the states  $|\varphi_g\>  |\varphi_{\gamma, g}\>$  (respectively, $|\varphi_g\> |\varphi_g\> |\varphi_{\gamma,g}\>$) belong to a single irreducible subspace, denoted by $\spc H_{\gamma_1}$  (respectively,  $\spc H_2$).  Precisely, they belong to the irreducible subspace that carries the  Cartan component of  $\hat U  \otimes \hat U^\gamma$  (respectively, $\hat U \otimes \hat U \otimes \hat U^\gamma$).
  By Schur's lemma the integral in the r.h.s. of both equations  (\ref{botheqs}) is proportional to a projector, namely
\begin{align}
\label{a}\tau_{\gamma}    &=   d_\gamma\Tr_{\gamma}  \left[   \frac{      \hat P_{ \gamma_1}}{d_{\gamma_1  }}  \left(      I \otimes |\varphi_\gamma\>\<\varphi_\gamma|  \right) \right]\\
\label{b}\rho_{\gamma}    &=  d_\gamma\Tr_{\gamma}  \left[ \frac{   \hat P_{ \gamma_2}}{d_{\gamma_2  }}  \left(   I  \otimes   I  \otimes |\varphi_\gamma\>\<\varphi_\gamma|  \right) \right]
\end{align}
where $\hat P_{\gamma_1}$ ($\hat P_{\gamma_2}$) is the projector on $\spc H_{\gamma_1}$ ($\spc H_{\gamma_2}$) and $d_{\gamma_1} $  ($d_{\gamma_2}$) is the formal dimension of the Cartan component of $\hat U  \otimes \hat U^\gamma$ ($\hat U \otimes \hat U \otimes \hat U^\gamma$).
Clearly, choosing $\lambda =   d_{\gamma_2}/d_{ \gamma_2} $, we have $  \lambda  \left(I  \otimes   \frac{  \hat P_{\gamma_1}}{d_{\gamma_1}}\right)\ge     \frac{ \hat P_{\gamma_2}}{d_{\gamma_2}} $, and, by Eqs. (\ref{a}) and (\ref{b}), $\lambda  (I\otimes \tau_\gamma)  \ge \rho_\gamma$.  Hence, we conclude that $ \fc (\gamma)  \le  d_{\gamma_1}/d_{\gamma_2}$.

On the other hand, the upper bound can be achieved.  First of all, note that $|\varphi\>$ is an eigenvector of $\tau_\gamma$: indeed, using Eq. (\ref{a}) one has
$$\tau_\gamma  |\varphi\>    =
   d_\gamma\Tr_{\gamma}  \left[   \frac{      \hat P_{\gamma_1}}{d_{\gamma_1  }}  \left(      I \otimes |\varphi_\gamma\>\<\varphi_\gamma|  \right) \right]  |\varphi\>=
   \frac{d_\gamma}{d_{\gamma_1}}    |\varphi\>.$$
Similarly,  $|\varphi\>|\varphi\>$ is an eigenvector of $\rho_\gamma$:  using Eq. (\ref{b}) one has
$$
\rho_\gamma  |\varphi\>  |\varphi\>    =
   d_\gamma\Tr_{\gamma}  \left[   \frac{      \hat P_{ \gamma_2}}{d_{\gamma_2  }}  \left(      I\otimes I \otimes |\varphi_\gamma\>\<\varphi_\gamma|  \right) \right] |\varphi\> |\varphi\>=
   \frac{d_\gamma}{d_{\gamma_2}}    |\varphi\> |\varphi\>.
$$
Hence, we obtain
\begin{align*}
\fc (\gamma)  & =    \left\|    (  I_  \otimes \tau_\gamma^{-1/2})  \rho_\gamma ( I\otimes \tau_\gamma^{-1/2} )  \right\|_\times \\
 & \ge    \< \varphi|  \<\varphi |    (  I  \otimes \tau_\gamma^{-1/2})  \rho_\gamma ( I\otimes \tau_\gamma^{-1/2} )   |\varphi\>|\varphi\> \\
 &  = \frac{d_{\gamma_2}}{d_{\gamma_2}}.
\end{align*}
Combining the upper and lower bounds we obtain  $\fc (\gamma)   =  {d_{\gamma_1}}/{d_{ \gamma_2}}$. Using Eq. (\ref{di}) for the evaluation of $d_{\gamma_1}$ and $d_{\gamma_2}$ concludes the proof. \qed

Using Eq. (\ref{probcft}) it is immediate to recover the benchmarks for coherent states and for squeezed states: indeed, one has
\begin{align}
\frac{  \int  \d^2 \alpha/\pi     ~\lambda  e^{-\lambda |\alpha|^2} ~   |\<  0  | \alpha\>  |^4  } {  \int  \d^2 \alpha/\pi     ~\lambda  e^{-\lambda |\alpha|^2} ~   |\<  0  | \alpha\>  |^2  }  = \frac{1+\lambda}{2+\lambda}  =  \fc^C (\lambda)
\end{align}
and
\begin{align}
\frac{  \int  \d s \d \theta/2\pi  ~  [\beta \sinh s/  (\cosh s )^{1+\beta}] ~   |\<  0  | \xi\>  |^4  }
{ \int  \d s \d \theta/2\pi   ~  [\beta \sinh s/ (\cosh s)^{1+\beta}]   ~   |\<  0  | \xi\>  |^2  }
 = \frac{1+\beta}{2+\beta}  =  \fc^S (\beta).
\end{align}
Note that for squeezed states the group-theoretical argument of Theorem \ref{theo:cft} guarantees optimality only for integer $\beta$ (the square-summable irreducible representations of $SU(1,1)$ form a discrete set), while the optimality proof for general real-valued positive $\beta$ requires the explicit argument presented in section \ref{sec:squeez}.

\section{Proof of Eq. (\element{3b}): benchmark for general pure Gaussian states}
Pure Gaussian states can be parameterized as
\begin{align*}
|\alpha, \xi\>  &:  =  \hat D(\alpha)  \hat S(\xi)    |0  \>\,,\qquad \xi = s e^{i\theta}.
\end{align*}
Here the displacement and squeezing operators generate a representation of the Jacobi group, and the vacuum is a lowest weight vector for this representation \cite{Sberceanu}.
Note that the overlap between one pure Gaussian state and the vacuum is
\begin{equation}\label{ove}
|\<  0  |  \alpha, \xi  \>  |^2  =     \frac{ e^{-  |\alpha|^2   +   {\rm Re} (  e^{-i\theta} \alpha^2)  \tanh s    } }{\cosh s}
\end{equation}
For the probability distribution, we choose
\begin{align*}
p^G_{\lambda,\beta}(  \alpha , s,\theta)   ~  \d^2\alpha \d s \d \theta  \propto    | \<  0   |  \lambda\alpha,\xi\>  |^2~     | \<  0   |  \xi\>  |^{2(4+\beta)} ~     \nu(\d^2 \alpha, \d^2\xi ),
\end{align*}
where $\nu (\d^2\alpha,\d^2 \xi) =   \d^2\alpha  \sinh s (\cosh s)^3 \d s \d \theta$ is the invariant measure over the Jacobi group.
Using Eq. (\ref{ove}), we can write down the explicit expression
\begin{eqnarray*}
\hspace*{-1cm}p_{\lambda,\beta}^G(\alpha,s,\theta)&=&\frac{\lambda \beta}{2 \pi^2}        \frac{ e^{-  \lambda|\alpha|^2   + \lambda   \tanh s      {\rm Re} (  e^{-i\theta} \alpha^2)}   \sinh s}{(\cosh s)^{\beta + 2}}  .
\end{eqnarray*}

For integer $\beta$, the states $  |\lambda\alpha,\xi\>  |\xi\>^{\otimes{4+\beta}}$ are Gilmore-Perelomov coherent states generated by the action of  the representation $  \hat D(\lambda \alpha)  \hat S(\xi)  \otimes \hat S(\xi)^{\otimes 4+\beta}$ on the lowest weight vector $|0\>   |0\>^{\otimes (4+\beta)}$.
Hence, using Eq. (\ref{probcft}) we can  compute the probabilistic CFT as
\begin{align*}
\fc^G(\lambda,\beta) &= \frac
{\int \d^2\alpha\,\d s \,\d\theta\ p^G_{\lambda,\beta}(\alpha,s,\theta)    | \<  0  |  \psi_{\alpha,s,\theta} \> |^4  }
{\int \d^2\alpha\,\d s \,\d\theta\ p^G_{\lambda,\beta}(\alpha,s,\theta)       | \<  0  |  \psi_{\alpha,s,\theta} \> |^2 }\\
& =    \frac     {  \int        \d^2\alpha\,  \d s\,  \d \theta\     e^{-  (2+ \lambda  ) \left[ |\alpha|^2   +   {\rm Re}  ( e^{-i\theta} \alpha^2)  \tanh s    \right]  }  ~ \sinh s    (\cosh s)^{-4 - \beta}  }
  {  \int        \d^2\alpha\, \d s\,  \d \theta\      e^{-  (1+ \lambda  ) \left[ |\alpha|^2   +   {\rm Re}  ( e^{-i\theta} \alpha^2)  \tanh s    \right]  }  ~ \sinh s    (\cosh s)^{-3 - \beta}  }  \\
&=\frac{\int\d^2\alpha\,\d s\ e^{-(2+\lambda) \left| \alpha \right| ^2} \sinh s  (\cosh s)^{-4-\beta}
   I_0\left((2+\lambda) \left| \alpha \right| ^2 \tanh s\right)}
   {\int\d^2\alpha\,\d s\  e^{-(1+\lambda) \left| \alpha \right| ^2}\sinh s  (\cosh s)^{-3-\beta}
   I_0\left((1+\lambda) \left| \alpha \right| ^2 \tanh s\right)}
   \\
  &=  \left( \frac{1+\lambda}{2+\lambda} \right)  \frac
{\int\d s\ \sinh s (\cosh s)^{-3-\beta}}
{\int\d s\ \sinh s (\cosh s)^{-2-\beta}} \\
&=  \frac{(1+\lambda)(1+\beta)}{(2+\lambda)(2+\beta)} \,.
\end{align*}
For general noninteger positive $\beta$ we prove the optimality of the value $ \frac{(1+\lambda)(1+\beta)}{(2+\lambda)(2+\beta)}$ directly from the expression of Ref.~\cite{Sgiulionew}, which in this case reads
\begin{align*}
 \fc^G (\lambda,\beta) & =   \left\| A_{\lambda ,\beta}    \right\|_\times   \qquad
 \left \{
 \begin{array}{ll} A_{\lambda,\beta} &: = (  I  \otimes \tau_{\lambda,\beta}^{-1/2})  \rho_{\lambda,\beta} ( I \otimes \tau_{\lambda,\beta}^{-1/2} )\\
   \tau_{\lambda, \beta}    &:=  \int  \d^2\alpha \d s\d\theta ~  p_{\lambda,\beta}  (\alpha, s,\theta)  ~   |\alpha,s,\theta\>\< \alpha,s,\theta|   \\
   \rho_{\lambda, \beta}    &:=  \int  \d^2\alpha \d s\d\theta ~  p_{\lambda,\beta}  (\alpha, s,\theta)  ~   |\alpha,s,\theta\>\< \alpha,s,\theta|\otimes    |\alpha,s,\theta\>\< \alpha,s,\theta|  \end{array}
   \right.
     \end{align*}
 In order to show that  $   \left\| A_{\lambda ,\beta}    \right\|_\times  =  \frac{(1+\lambda)(1+\beta)}{(2+\lambda)(2+\beta)}$, we observe that, by the symmetry of the prior distribution,   the average state $\tau_{\lambda,\beta}$ commutes with the phase shifts $\hat U_\varphi  =  e^{i\varphi \hat a^\dag a}$, and, therefore, it is diagonal on the Fock basis. Hence, we can write it as
\begin{equation*}
\tau_{\lambda,\beta}  =  \sum_{n=0}^{\infty}   t_n |n\>\<n|\,,  ~\qquad t_n =  \int  \d^2 \alpha ~\d s~ \d \theta ~  p_{\lambda,\beta}^G(\alpha,s,\theta)  ~   | \<n  |   \hat D(\alpha)  \hat S(\xi)  |0\>|^2
\end{equation*}
In particular, the vacuum $|0\>$ is eigenvector of $\tau_{\lambda,\beta}$ for the eigenvalue $t_0 =  \int  \d^2 \alpha ~\d s~ \d \theta ~  p_{\lambda,\beta}^G(\alpha,s,\theta)  ~   | \<0  |   \hat D(\alpha)  \hat S(\xi)  |0\>|^2$.
Likewise, the state $\rho_{\lambda,\beta}$ commutes with the phase shifts $\hat U_\varphi \otimes \hat U_{\varphi}$, and, therefore, can be diagonalized jointly with the total number operator $\hat a_1^\dag \hat a_1 + \hat a_2^\dag \hat a_2$  ($\hat a_1$ and $\hat a_2$ representing the annihilation operators on the two copies of the Hilbert space).
This implies that  $|0\>  |0\>$ is an eigenvalue of $\rho_{\lambda,\beta}$, with eigenvalue $r_{00} =  \int  \d^2 \alpha ~\d s~ \d \theta ~  p_{\lambda,\beta}^G(\alpha,s,\theta)  ~   | \<0  |   \hat D(\alpha)  \hat S(\xi)  |0\>|^4$.

Combining these observations, we obtain the lower bound
\begin{align}
\nonumber \fc^G (\lambda,\beta) & =  \|  A_{\lambda,\beta}\|_\times\\
\nonumber  &= \left\|  (  I  \otimes \tau_{\lambda,\beta}^{-1/2})  \rho_{\lambda,\beta} ( I \otimes \tau_{\lambda,\beta}^{-1/2} )   \right\|_\times \\
\nonumber  &  \ge   \< 0  | \<  0 |(  I  \otimes \tau_{\lambda,\beta}^{-1/2})  \rho_{\lambda,\beta} ( I \otimes \tau_{\lambda,\beta}^{-1/2} )|0\>|0\>\\
\nonumber  &  =  r_{00}/t_0\\
&  =   \frac{(1+\lambda)(1+\beta)}{(2+\lambda)(2+\beta)}.\label{low}
\end{align}
On the other hand, the  explicit evaluation of all the eigenvalues of $A_{\lambda,\beta}$ can be carried out by diagonalizing the submatrices  $A^{(k)}_{\lambda,\beta}$ corresponding to the compression of $A_{\lambda,\beta}$ onto the subspaces with total photon number $k$. The matrix elements of $A^{(k)}_{\lambda,\beta}$ are  given by
\begin{equation}\label{eqei}
\left[  A^{(k)}_{\lambda,\beta}\right]_{mn}   =   \int  \d^2 \alpha ~\d s~ \d \theta ~  p_{\lambda,\beta}^G(\alpha,s,\theta)  ~   \frac{  \<n  |   \hat D(\alpha)  \hat S(\xi)  |0\>}{\sqrt{t_n}}   \<k-n  |   \hat D(\alpha)  \hat S(\xi)  |0\>      \frac{ \overline{ \<m  |   \hat D(\alpha)  \hat S(\xi)  |0\>}}{\sqrt{t_m}}  \overline{ \<k-m  |   \hat D(\alpha)  \hat S(\xi)  |0\>   }\,,
\end{equation}
for $m,n  =  0,\dots, k.$
Compact expressions for the overlap  $\<n  |   \hat D(\alpha)  \hat S(\xi)  |0\>$, involving Hermite polynomials, can be found in \cite{Snortonetal}. The explicit calculation of the eigenvalues of $A^{(k)}_{\lambda,\beta}$, done by numerical methods, confirms that $ \left[ A_{\lambda,\beta}^{(0)}\right]_{00}$ is the maximum eigenvalue.  This gives the upper bound   $\| A_{\lambda,\beta}\|_\times \le \frac{(1+\lambda)(1+\beta)}{(2+\lambda)(2+\beta)}$, which, combined with the lower bound of Eq.~(\ref{low}), leads to  the benchmark $\fc^G (\lambda, \beta) = \frac{(1+\lambda)(1+\beta)}{(2+\lambda)(2+\beta)}$ for all $\lambda,\beta>0$.

\section{Prior distributions and optimal measurements for the estimation of Gilmore-Perelomov coherent states}
Consider  the probability distribution $p(g) \d g =  d    |\<\varphi|\varphi_g\> |^2  \d g $ defined by the Gimore-Perelomov coherent states $|\varphi_g\> =   \hat U_g |\varphi\>$ associated to a square-summable representation  $\hat U  :  g \mapsto \hat U_g$.    This probability distribution can be generated  by   preparing a quantum system in the state  $|\varphi\>$ and by performing the quantum measurement with POVM
\begin{align}\label{povm}
\hat P_g~ \d g  =    d    |\varphi_g\>\<  \varphi_g|\,,  \qquad d  =   \left( \int \d g ~     |\<\varphi|\varphi_g\> |^2  \right)^{-1} \, ,
\end{align}
whose integral is normalized to the identity thanks to Schur's lemma.

We now show that the POVM $\{\hat P_g\}_{g\in\grp G}$ is the optimal measurement for the estimation of the parameter $g$ characterizing the coherent state $|\varphi_g\>$.
We denote by $\check g$ the estimated  value, and by $g$ the true one and define a figure of merit that assigns a score to the estimate depending on how close it is to the true value.  As a figure of merit, we consider here the estimation fidelity $f (\check g, g)  =  |  \< \varphi_{\check g} |  \varphi_g\> |^{2}$.
With these settings, the average score achieved by a generic POVM $\hat Q_g \d g $ is   ${\cal F}_g  =  \int \d \check g~   f(\check g ,g)   \<\varphi_g| \hat Q_{\check g}   |\varphi_g\> $.

Here we do not assume any prior probability on $g$ and consider instead  the maximization of the worst-case fidelity  ${\cal F}_{wc}  =  \inf_{g\in\grp G}  {\cal F}_g$.
For this problem, it is known that the optimal POVM can be found in the set of covariant POVMs \cite{Shelstrom,Sholevo,Sozawa}, which in our case have the form  $\hat Q_g  = d~  \hat U_g  \omega  \hat U_g^\dag$, where $\omega$ is a density matrix. For a covariant POVM, the fidelity is independent of $g$, and therefore one has ${\cal F}_{wc}  =  {\cal F}_g ~\forall g\in\grp G$ Using this fact, it is easy to prove the following
\begin{Theo}[Optimal estimation of Gilmore-Perelomov coherent states]
The maximum of the worst-case fidelity for  the estimation of a Gilmore-Perelomov coherent state $|\varphi_g\>$ is
\begin{align}
{\cal F}_{wc}  =   \frac{  \int \d g ~  |  \< \varphi | \varphi_g\>|^4}{  \int \d g ~  |  \< \varphi | \varphi_g\>|^2}\, .
\end{align}
The optimal POVM achieving the maximum fidelity is given by Eq. (\ref{povm}).
\end{Theo}
\Proof
Choosing the true value to be the identity element in the group ($g  =  e$), the fidelity for a generic POVM $\hat Q_g  = d~ \hat  U_g  \omega  \hat U_g^\dag$ can be evaluated as
\begin{align*}
{\cal F}_{wc}   [\hat Q_g \d g]  & =  d~ \int \d \check g~      f(\check g  |  e)   ~  \< \varphi |  \hat U_{\check g} \omega   \hat U_{\check g}^\dag |\varphi\>  \\
& = d~    \Tr  \left[ \left( |\varphi\>\<\varphi|  \otimes \omega \right) \left(    \int \d \check g~    |\varphi_g\>\<\varphi_g|    \otimes  |\varphi_g\>\<\varphi_g|  \right) \right ]  \\
&  =\frac{d}{d'}     \Tr  \left[ \left( |\varphi\>\<\varphi|  \otimes \omega \right)  P' \right ]
\end{align*}
where $P'$ is the projector on the irreducible subspace  $\spc H'$  spanned by the coherent states $\{|\varphi_g\>|\varphi_g\>\}_{g\in\grp G}$ and $d'$ is the corresponding formal dimension, $d' =    \left( \int \d g ~     |\<\varphi|\varphi_g\> |^4  \right)^{-1} $.
Hence, we conclude that ${\cal F}_{wc}   [\hat Q_g \d g]    \le d/d'$.  The upper bound can be achieved by choosing the POVM $\hat P_g $. Recalling the definitions of $d$ and $d'$, we then have the desired result. \qed

\medskip

In the specific case of the group $SU(1,1)$, the optimality of the POVM of Eq. (\ref{povm}) was derived by Hayashi \cite{Shayashi}.   Note also that the above proof applies also to different figures of merit, provided that they are of the form $  f_\gamma(\check g, g)  =   |\<  \varphi_\gamma|\varphi_{\gamma ,g}\>|^2$, for some Gilmore-Perelomov coherent state $|\varphi_{\gamma, g}\>  = \hat  U^\gamma_{g}  |\varphi_\gamma\>$ associated to some irreducible representation $\hat U^\gamma:  g  \mapsto \hat U_g^
\gamma$.  In this case, Eq.~(\ref{povm}) still gives the optimal POVM and the optimal fidelity reads
\begin{align}
{\cal F}_{wc}  =   \frac{  \int \d g ~    |\<  \varphi_\gamma|\varphi_{\gamma,g} \>|^2 |  \< \varphi | \varphi_g\>|^2}{  \int \d g ~  |  \< \varphi | \varphi_g\>|^2}\, .
\end{align}

\clearpage
\end{widetext}

\end{document}